\documentclass{elsarticle}

\usepackage{amssymb}
\usepackage{amsmath}
\usepackage{stmaryrd}
\usepackage{ulem}
\usepackage{esint}
\usepackage{graphicx}
\usepackage{caption}
\usepackage{subcaption}
\usepackage{xfrac}
\usepackage{diagbox}
\usepackage{xcolor}
\usepackage{natbib}
\usepackage{floatrow}

\newfloatcommand{capbtabbox}{table}[][\FBwidth]

\graphicspath{{figures/}}

\journal{Journal of Computational Physics}

\begin{document}

\begin{frontmatter}

\title{A Conservative Cartesian Cut Cell Method for the Solution of the Incompressible Navier-Stokes Equations on Staggered Meshes}

\author[ijlra]{Alejandro Quir{\'o}s Rodr{\'i}guez}
\ead{alejandro.quiros_rodriguez@sorbonne-universite.fr}

\author[ijlra]{Tomas Fullana}
\ead{tomas.fullana@sorbonne-universite.fr}

\author[msme]{Vincent Le Chenadec\corref{corresponding}}
\cortext[corresponding]{Corresponding author}
\ead{vincent.le-chenadec@univ-eiffel.fr}

\author[ijlra,ITV]{Taraneh Sayadi}
\ead{taraneh.sayadi@sorbonne-universite.fr}

\address[ijlra]{Sorbonne Universit{\'e}, Institut Jean le Rond d'Alembert, IJLRA, F-75005 Paris, France}
\address[msme]{MSME, Universit\'{e} Gustave Eiffel, UPEC, CNRS, F-77454 Marne-la-Vall{\'e}e, France}
\address[ITV]{Institute of Combustion Technologies, RWTH-Aachen University, Aachen, Germany}

\begin{abstract}
The treatment of complex geometries in Computational Fluid Dynamics applications is a challenging endeavor, which immersed boundary and cut-cell techniques can significantly simplify by alleviating the meshing process required by body-fitted meshes. These methods however introduce new challenges, as the formulation of accurate and well-posed discrete operators becomes nontrivial. Here, a conservative cartesian cut cell method is proposed for the solution of the incompressible Navier--Stokes equation on staggered Cartesian grids. Emphasis is set on the structure of the discrete operators, designed to mimic the properties of the continuous ones while retaining a nearest-neighbor stencil. For convective transport, a divergence is proposed and shown to also be skew-symmetric as long as the divergence-free condition is satisfied, ensuring mass, momentum and kinetic energy conservation (the latter in the inviscid limit). For viscous transport, conservative and symmetric operators are proposed for Dirichlet boundary conditions. Symmetry ensures the existence of a sink term (viscous dissipation) in the discrete kinetic energy budget, which is beneficial for stability. The cut-cell discretization possesses the much desired summation-by-parts (SBP) properties. In addition, it is fully conservative, mathematically provably stable and supports arbitrary geometries. The accuracy and robustness of the method are then demonstrated with flows past a circular cylinder and an airfoil.
\end{abstract}


\end{frontmatter}


\section{Introduction}

A vast range of flow phenomena are dominated by the dynamics that occur within the vicinity of solid boundaries. These include the viscous and pressure drag observed in external flows, the conjugate heat transfer blades are subjected to in gas turbines or the generation of vorticity in boundary layers and its subsequent impact on the turbulent mixing, to name a few examples. The effect of the dynamics in the vicinity of boundaries on the overall flow singles out the treatment of boundary conditions, a critical aspect that also represents a significant challenge for many Computational Fluid Dynamics (CFD) applications.

Many numerical methods have therefore been developed to address the treatment of boundary conditions on complex geometries. Unstructured techniques, as the name suggests, leverage meshes with arbitrary polyhedral elements that at least for piece-wise planar cases conform to the geometry, at the cost of explicitly storing connectivity information.
They are very effective and powerful to represent arbitrary geometries, and can even represent curved surface exactly~\cite{Hughes2005}, but the generation of high-quality unstructured meshes continues to be a challenging and time-consuming task. In addition, the design of efficient and robust numerical algorithms targeting such meshes remains an active area of research~\cite{Mavriplis1995}. Finally, explicit element connectivity effectively introduces an overhead that does not exist on structured meshes, and consequently increases the computational cost per grid point.


These limitations are one of two compelling arguments for the use of structured meshes, the second being the simplicity and efficiency of the implementation of many algorithms on such meshes. The connectivity is implicit, which restricts their use to simple mesh topologies, including cylindrical or curvilinear ones. To circumvent this limitation, dedicated discretization techniques, referred to as immersed boundary methods (IBM), have been devised~\cite{Mittal2005}. There exists various approaches to represent the boundary (diffuse or sharp) and to account for the mass and momentum transfers that occur along the solid boundary. The original IBM~\cite{Peskin1972}, which targeted cardiovascular flows, represented the boundary as a flexible elastic membrane, which enabled the explicit expression of the force exerted onto the flow. This approach however is not valid for rigid boundaries, for which a variety of techniques ranging from the use of fictitious domain methods~\cite{Glowinski1994} and Lagrange multiplier methods~\cite{Taira2007}.

IBM techniques have been adapted to suit the numerical representation of PDE solutions, such as the Finite Difference Method and the Finite Element Method. A widespread flavor of the IBM, favored by the Finite Volume community, is referred to as the cut-cell method, developed for scalar equations~\cite{Calhoun2000} or the viscous compressible flows~\cite{Berger2012a,Schneiders2016} on collocated Cartesian grids, and for incompressible Navier-Stokes equations on staggered Cartesian grids~\cite{Cheny2010}. The combination of the cut cell method with staggered arrangement (also referred to as Arakawa C grid~\cite{Arakawa1977}), adopted in two dimensions by Cheny and Botella~\cite{Cheny2010}, is a sensible choice for incompressible flows: it guarantees a strong coupling between the pressure and velocity variables~\cite{Harlow1965}, and can potentially conserve important physical invariants such as kinetic energy in the inviscid limit of the incompressible Navier-Stokes equations~\cite{Morinishi1998}. Preserving such properties in the presence of complex boundaries is however a challenge which, to the best of the authors' knowledge, is yet to be fulfilled.

The proposed method attempts to fill this void. The formulation, delineated in the following section, is flexible enough to support geometry defined by various means, such as Constructive Solid Geometry primitives or surface triangulations, provided a finite set of geometric moments can be computed from them, such as the centroid coordinates of wet volumes or the area of wet faces. One advantage of this method is that these geometric fields are the only information required to modify classical finite differences formulas in the vicinity of boundaries. Well-known second-order formulas are also recovered away from the boundaries , and the formulations accommodates any stretching. The definition of these geometric fields and their number is determined from accuracy considerations. It will be shown in particular that the proposed operators degenerates to classical formulas for the mesh-aligned boundaries.

The discrete calculus of Morinishi~\cite{Morinishi1998,Morinishi2010} is leveraged to provide concise expressions for the discrete operators, for Dirichlet boundary conditions imposed on the velocity field. In addition, the expressivity of Morinishi's calculus allows for a systematic analysis of the structure of the pressure gradient, velocity divergence as well as convective and viscous transport operators. First, all operators are shown to preserve constant states, in the boundary vicinity or away from it (free-streaming conditions). Second, divergence, advective and skew-symmetric versions of the convective transport are proposed and shown to be equivalent and both momentum- and kinetic-energy-conserving upon satisfaction of the continuity equation (divergence-free condition). Third, a Dirichlet version of the viscous transport is proposed and shown to be symmetric positive definite, which results in the dissipation term in the discrete kinetic energy equation to be positive all the way to the boundary for viscous flows. Standard validations are provided that assess the scheme's accuracy and stability.

The manuscript is structured as follows. Sec. 2 motivates the choice made in the design of the method, in particular the set of geometric moments that must be computed from the geometry. Sec. 3 precisely defines these moments, and the set of notations used through Sec. 4 which introduces the semi-discretisation as well as the segregated approach used for time-integration of the incompressible Navier-Stokes equations for a Newtonian fluid. Sec. 5 presents the flow solution around a cylinder and an airfoil and compares them to reference solutions.

\section{Motivation} \label{sec:methodology}

This section motivates the choices underlying the design of the proposed cut-cell operators. To do so, the focus is set on the numerical solution of the Poisson problem
$$
\Delta T = \sigma
\label{eq:laplacian}
$$
where $\sigma$ is a specified source term and $T$ is also subject to a Dirichlet boundary condition $D$. Simply put, the question addressed here is: what is the minimal amount of geometric information required to discretize the Poisson equation on an arbitrary domain using Cartesian grid, while guaranteeing that the discrete Laplacian operator (i) preserves a classical three-point star-shaped stencil, while (ii) guaranteeing first order accuracy in mesh-aligned cases.
The construction of this operator will ultimately serve for the discretization of the viscous term in the incompressible Navier-Stokes equations.

\subsection{Governing principles}
Cut Cell Methods are firmly grounded in the Finite Volume Method, which defines the primary discrete variables as cell-wise averages over mesh elements (as opposed to point-wise values in the Finite Difference Method, for example). The design of the Finite Volume operators is then based on the application of Stokes' theorem. For example, given a scalar field $T$,  this theorem states that in a Cartesian coordinate system, the $x$ component of the gradient $\mathbf{q} \equiv \nabla T$ averaged over a cell $\Omega$ may be computed as
\begin{equation}
\left \vert \Omega \right \vert q _ x = \int _ \Omega \frac{\partial T}{\partial x} \mathrm{d} V = \oint _ {\partial \Omega} T \mathbf{e} _ x \cdot \mathrm{d} \mathbf{S}
\label{eq:Stokes}
\end{equation}
where $\left \vert \cdot \right \vert$ denotes the measure operator, $\mathrm{d} \mathbf{S}$ the outward-pointing surface element, $\mathbf{e} _ x$ the unit vector along the $x$ direction and $\partial \cdot$ the contour operator.

\begin{figure}
    \centering
    \begin{subfigure}[b]{0.45\textwidth}
         \centering
         \includegraphics[width=\textwidth]{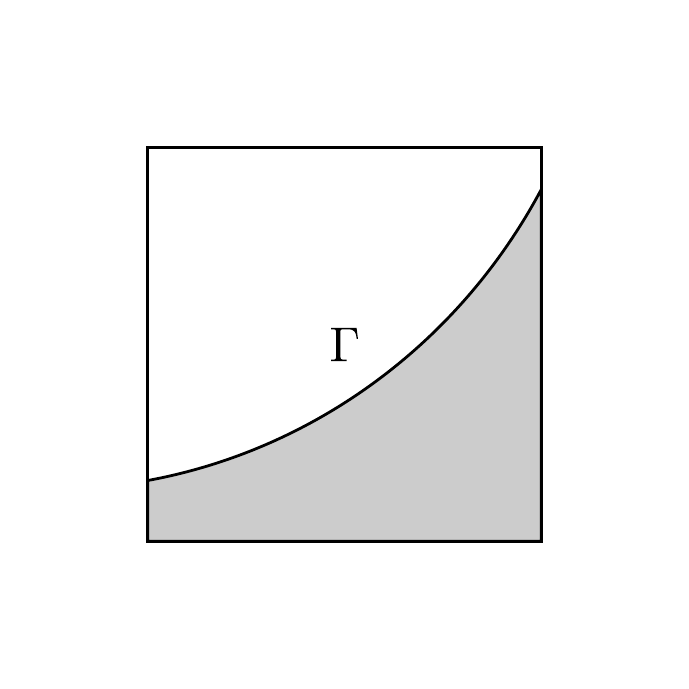}
         \caption{Exact}
         \label{fig:exact}
     \end{subfigure}
     \hfill
     \begin{subfigure}[b]{0.45\textwidth}
         \centering
         \includegraphics[width=\textwidth]{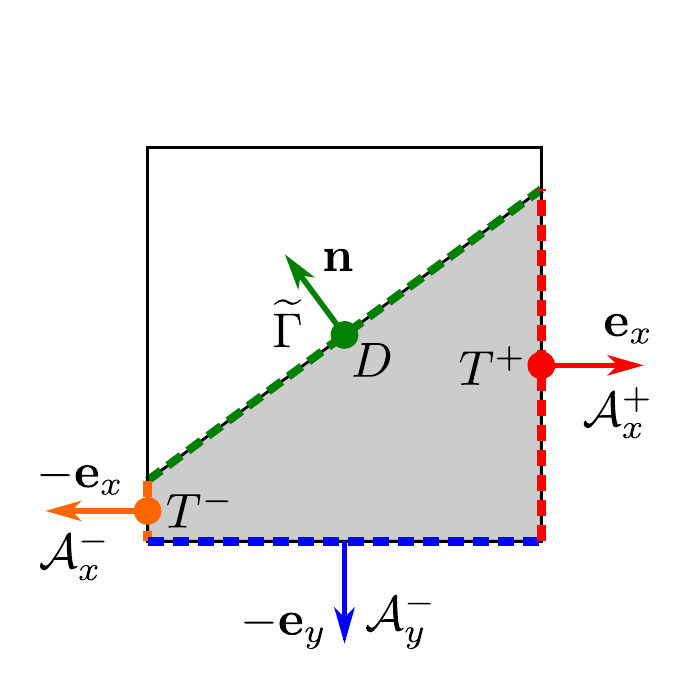}
         \caption{Approximate}
         \label{fig:approximate}
     \end{subfigure}
    \caption{Caption}
\end{figure}
For the sake of presentation, the case displayed in Fig.~\ref{fig:exact} is considered where $\Omega$ consists of the intersection of a phase domain and a computational cell (a right hexahedron). The contour $\partial \Omega$ then consists of the union of the three planar faces $\mathcal{A} _ x ^ -$, $\mathcal{A} _ x ^ -$ and $\mathcal{A} _ y ^ -$ as well as the boundary surface $\Gamma$. A piece-wise linear approximation of $\Gamma$, denoted $\widetilde{\Gamma}$, of length $\left \vert \widetilde{\Gamma} \right \vert$ and unit normal $\left ( n _ x, n _ y \right )$, can be defined as done in Fig.~\ref{fig:approximate}. Applying Eq.~\ref{eq:Stokes} to $\widetilde{\Omega}$ with $T = 1$ then yields
$$
\int _ {\widetilde{\Omega}} \frac{\partial 1}{\partial x} \mathrm{d} V = \left \vert \mathcal{A} _ x ^ + \right \vert - \left \vert \mathcal{A} _ x ^ - \right \vert + n _ x \left \vert \widetilde{\Gamma} \right \vert = 0
$$
which highlights the existence of a fundamental relation
\begin{equation}
\left \vert \mathcal{A} _ x ^ + \right \vert - \left \vert \mathcal{A} _ x ^ - \right \vert = -n _ x \left \vert \widetilde{\Gamma} \right \vert
\label{eq:approximate}
\end{equation}
sometimes referred to as a Surface Conservation Law (SCL).

In other words, the knowledge of $\left ( \left \vert \mathcal{A} _ \alpha \right \vert \right ) _ {\alpha \in \left \{ x, y, z \right \}}$ implicitly defines a piece-wise linear approximation to the boundary. As a consequence, this surface information, henceforth referred to as the surface capacity, may serve to approximate the right-hand side of Eq.~\ref{eq:Stokes}. If the unknowns $\left ( T ^ \pm _ {x/y} \right )$ are defined as averages over the wet areas $\left ( \left \vert \mathcal{A} ^ \pm _ {x/y} \right \vert \right )$, the formula
$$
\oint _ {\partial \widetilde{\Omega}} T \mathbf{e} _ x \cdot \mathrm{d} \mathbf{S} = \left \vert \mathcal{A} _ x ^ + \right \vert T ^ + - \left \vert \mathcal{A} _ x ^ - \right \vert T ^ - - \left ( \left \vert \mathcal{A} _ x ^ + \right \vert - \left \vert \mathcal{A} _ x ^ - \right \vert \right ) D
\label{eq:quadrature}
$$
is exact, provided $D$ is the Dirichlet condition averaged over the approximate boundary $\widetilde{\Gamma}$.

To complete the definition of the averaged $x$-component of the gradient, the volume capacity $\mathcal{V} \equiv \left \vert \Omega \right \vert$ is also required, which results in the following tentative gradient operator 
$$
Q _ x ^ {\mathrm{v} 1} \simeq \left ( \left \vert \mathcal{A} _ x ^ + \right \vert T ^ + - \left \vert \mathcal{A} _ x ^ - \right \vert T ^ - - \left ( \left \vert \mathcal{A} _ x ^ + \right \vert - \left \vert \mathcal{A} _ x ^ - \right \vert \right ) D \right ) / \mathcal{V}.
$$
It is worth stressing that the use of the SCL (Eq.~\ref{eq:approximate}) in $Q _ x ^ {\mathrm{v} 1}$ guarantees that the discrete gradient vanishes when the solution and boundary values are matching constants ($T ^ + = T ^ - = D$).

This notation can be generalized to arbitrary dimensions for any boundary geometry using the differentiation operator $\delta \cdot / \delta \xi _ \alpha$, $\alpha \in \left \{ x, y \right \}$ as follows
\begin{equation}
\forall \alpha \in \left \{ x, y \right \}, \quad \operatorname{grad} _ \alpha ^ {\mathrm{v}1} \left ( T _ \alpha, D \right ) = \frac{1}{V} \left ( \frac{\delta A _ \alpha T _ \alpha}{\delta \xi _ \alpha} - \frac{\delta A _ \alpha}{\delta \xi _ \alpha} D \right )
\label{eq:simplegradient}
\end{equation}
where all components of the discrete vector field $\uline{Q} = \left ( Q _ \alpha \right ) = \left ( \operatorname{grad} _ \alpha ^ {\mathrm{v}1} \left ( T _ \alpha, D \right ) \right )$ are collocated with $D$. In Eq.~\ref{eq:simplegradient}, the operator $\delta \phi \cdot / \delta \xi _ \alpha$ denotes the discrete differentiation operator along direction $\alpha$ on a mesh with unit spacing. When $\alpha = x$ and $\phi _ {i, j}$ is centered at $\left ( x _ i, y _ j \right )$, it is defined as
\begin{equation}
\left . \frac{\delta \phi}{\delta \xi _ x} \right \vert _ {i + \sfrac{1}{2}, j} = \phi _ {i + 1, j} - \phi _ {i j}.
\label{eq:differentiation}
\end{equation}
This definition is straightforward to extend to either staggered $( \phi _ {i + \sfrac{1}{2}, j} )$ and $( \phi _ {i, j + \sfrac{1}{2}} )$ or nodal $( \Phi _ {i + \sfrac{1}{2}, j + \sfrac{1}{2}} )$ fields. Likewise, differentiation in the second direction, $\delta ( \cdot ) / \delta \xi _ y$, is defined in the same manner. Finally, extension to three dimensions and restriction to one are obtained by adding and removing an index, respectively.

%


\begin{figure}
     \centering
     \begin{subfigure}[b]{0.45\textwidth}
         \centering
         \includegraphics[width=\textwidth]{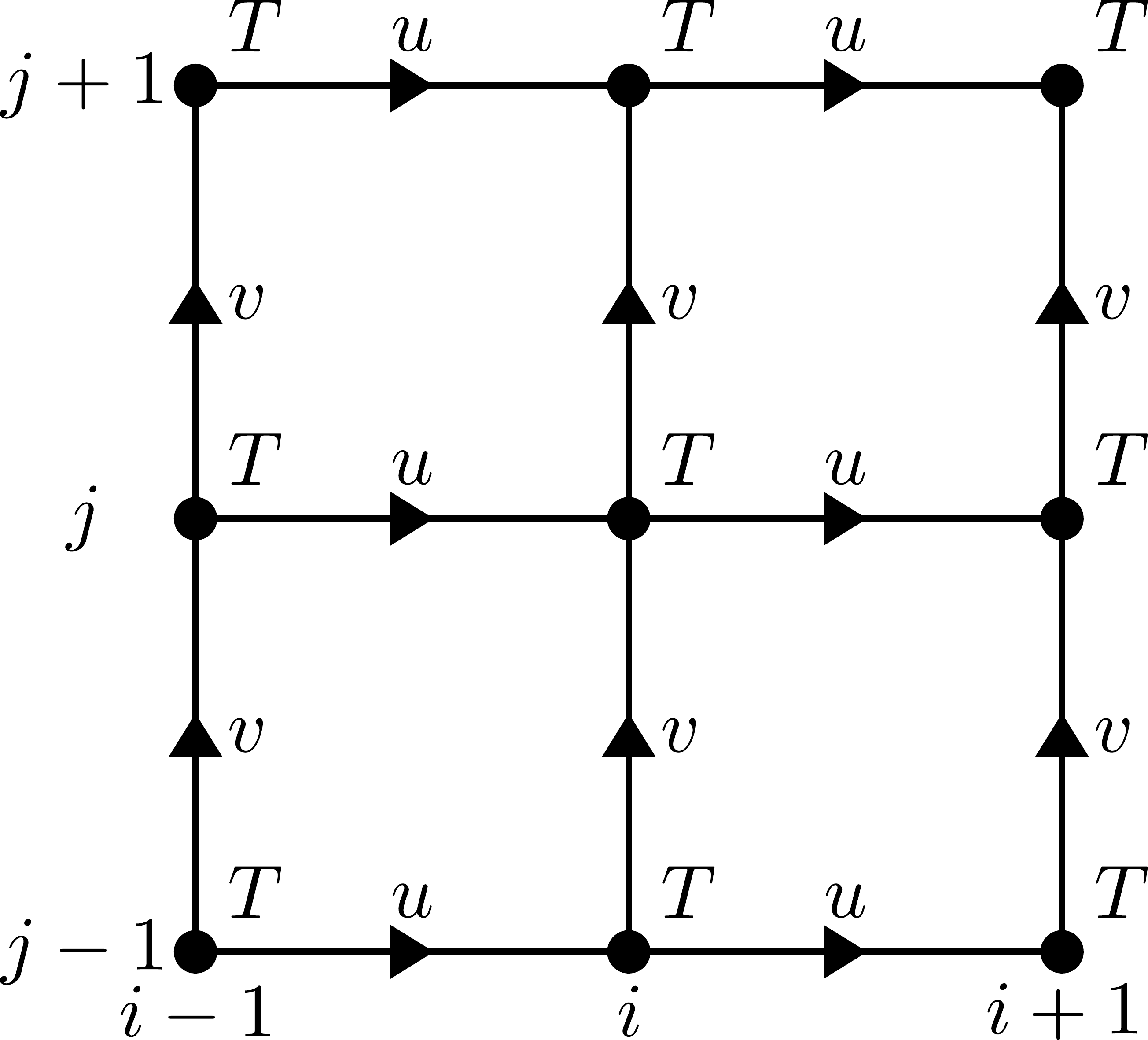}
         \caption{Arakawa C grid.}
     \end{subfigure}
     \hfill
     \begin{subfigure}[b]{0.47\textwidth}
         \centering
         \includegraphics[width=\textwidth]{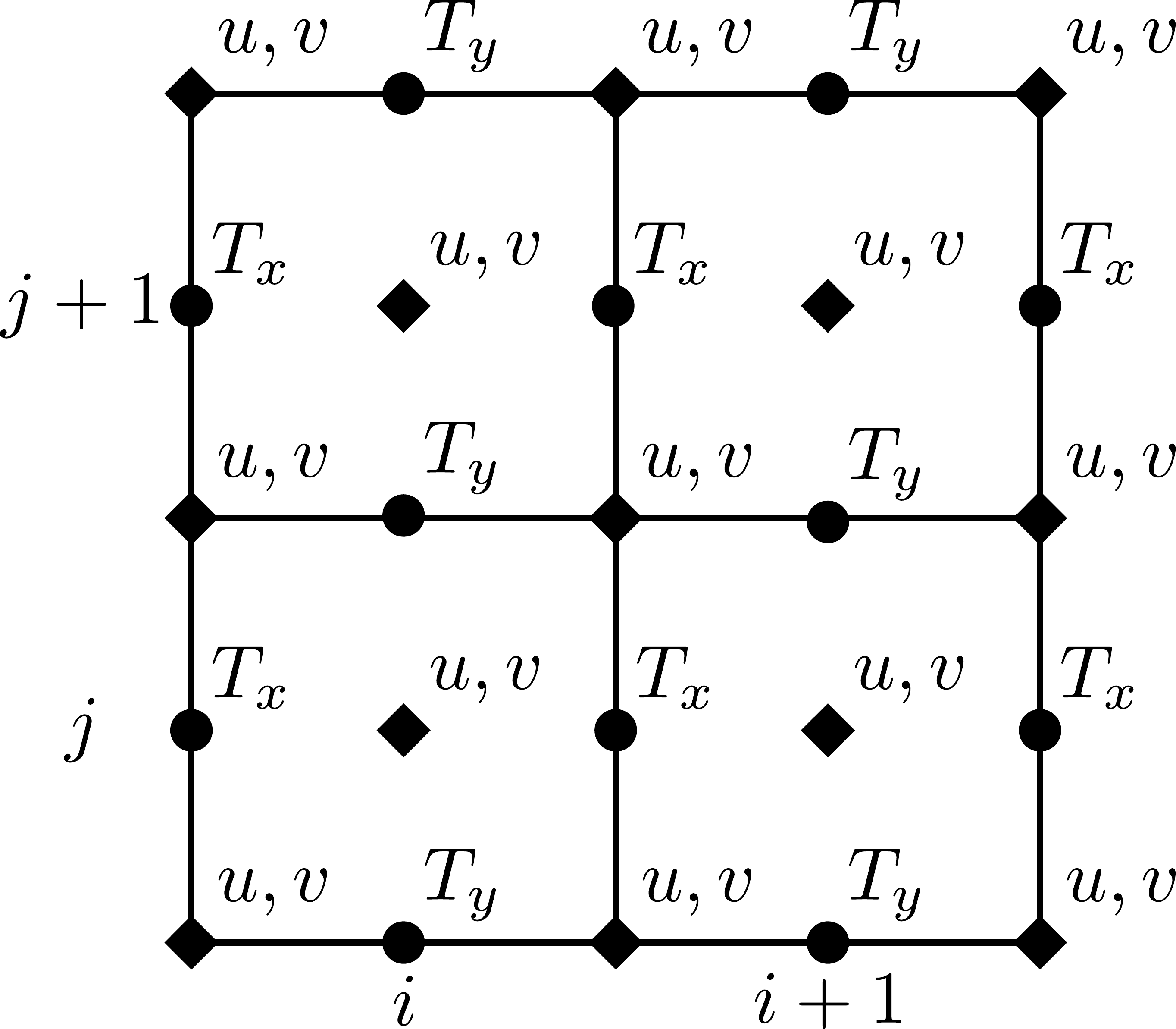}
         \caption{Arakawa E grid.}
     \end{subfigure}
     \caption{Variable arrangements considered in presented work.}
     \label{fig:arakawa}
\end{figure}

In this first version of the gradient operator, the primary unknowns $T _ \alpha$ are collocated with the surface areas $A _ \alpha$, whereas the Dirichlet boundary condition $D$ is staggered in between. This construction, referred to as Arakawa E grid~\cite{Arakawa1977} (see Fig.~\ref{fig:arakawa}), relies on the definition of multiple temperature fields. Such a grid configuration is not the one adopted by the MAC approach~\cite{Harlow1965}, which favors the C-grid that defines a single temperature field collocated with the $D$ field here. A C-grid however means that the temperature unknowns $T$ and surface capacities $A _ \alpha$ are staggered, in which case the latter together with $V$ should be interpolated as follows
$$
\forall \alpha \in \left \{ x, y \right \}, \quad \operatorname{grad} ^ {\mathrm{v}2} _ \alpha \left ( T, D \right ) = \frac{1}{\overline{V} ^ \alpha} \left ( \frac{\delta \overline{A} ^ \alpha T}{\delta \xi _ \alpha} - \overline{\frac{\delta A _ \alpha}{\delta \xi _ \alpha} D} ^ \alpha \right )
\label{eq:grad-v2}
$$
which introduces the interpolation operator $\overline{\cdot} ^ \alpha$, $\alpha \in \left \{ x, y\right \}$, defined in direction $x$ as
\begin{equation}
\left . \overline{\phi} ^ x \right \vert _ {i + \sfrac{1}{2}, j} = \frac{\phi _ {i + 1, j} + \phi _ {i j}}{2}
\label{eq:interpolation}
\end{equation}
for a field $\phi$ centered at $\left ( x _ i, y _ j \right )$. Interpolation in direction $y$ as well as extensions to staggered variables, are defined analogously, as previously discussed for differentiation operations.

\subsection{Loss of accuracy with interpolation} \label{sec:bay}

It should be noted that formulas other than Eq.~\ref{eq:grad-v2} can also be written without interpolation of the geometric capacities, for example by collocating all surface capacities $\left ( A _ \alpha \right )$ with the primary variable $T$. However in the context of a second order operator such as the scalar Laplacian (Eq.~\ref{eq:laplacian}), the need for interpolation will resurface in the approximation of the divergence operator. This section therefore focuses on the limitations of the second tentative formula (Eq.~\ref{eq:grad-v2}), more specifically its failure to revert to a classical first order approximation of the second order derivative in the limit where the boundary is orthogonal to the direction of interest. This is the central point of the proposed cut-cell method, namely the enhancement of the geometric description of the boundary by means of additional volume and surface capacities to revise the gradient and divergence operators so as to achieve first-order accuracy in the vicinity of mesh-aligned boundaries.



\begin{figure}
    \centering
    \includegraphics[width=\textwidth]{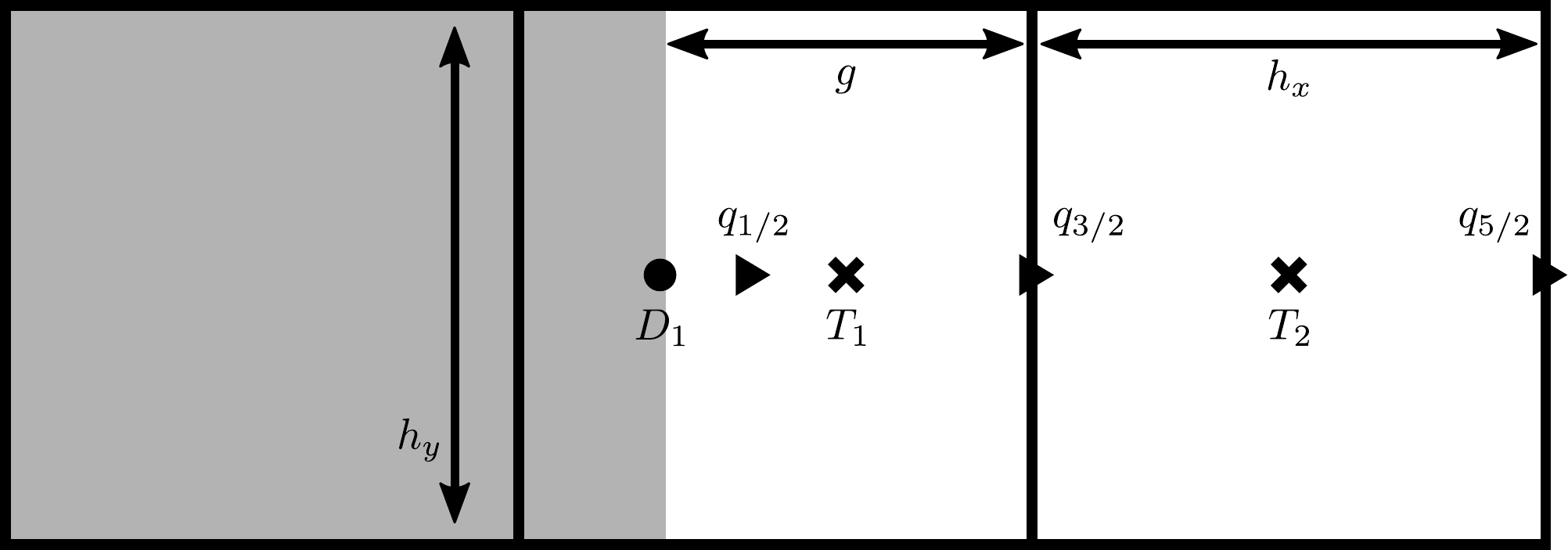}
    \caption{Insufficient geometric information resulting in loss of accuracy in mesh-aligned geometries.}
    \label{fig:interpolation}
\end{figure}

To illustrate the limitation of the tentative gradient formula (Eq.~\ref{eq:grad-v2}), the discretization of the second-order derivative along $x$ in the mesh-aligned two-dimensional configuration displayed in Fig.~\ref{fig:interpolation} is considered, where the fluid occupies the rightmost cells. This configuration is characterized by $V _ 0 = 0$, $V _ 1 = ( h _ x - g ) h _ y$, $V _ 2 = h _ x h _ y$, $A _ {1/2} = 0$ and $A _ {3/2} = A _ {5/2} = h _ y$ (here, $A$ stands for $A _ x$ since only the $x$ contribution is considered). Using these expressions, Eq.~\ref{eq:grad-v2} simplifies to $Q _ {-1/2} = 0$,
$$
Q _ {1/2} = \frac{T _ 1 - D _ 1}{g},
\label{eq:grad-v2-boundary}
$$
and
$$
Q _ {3/2} = \frac{T _ 2 - \left ( T _ 1 + D _ 1 \right ) / 2}{\left ( g + h _ x \right ) / 2}.
\label{eq:grad-v2-away}
$$
This approximation of the gradient is problematic for two reasons: (i) At the boundary, the $x$-gradient value ($Q _ {1/2}$) is under predicted by a factor of $2$, since the denominator of the right-hand side of Eq.~\ref{eq:grad-v2-boundary} stands at $g$ when it should match the distance between the points where $D _ 1$ and $T _ 1$ are defined, $g / 2$. (ii) Away from the boundary, the $x$-gradient value ($Q _ {3/2}$) depends on the boundary condition $D _ 1$, when one would simply expect to difference $T _ 2 - T _ 1$ to appear in the numerator of the right-hand side of Eq.~\ref{eq:grad-v2-away}.



This simple exercise highlights the loss of accuracy associated with the interpolation of the geometric capacities. This can be associated with the fact that they are defined as volume and surface integrals of the characteristic function of the fluid domain $\Omega ^ f \subset \Omega$, defined as
\begin{equation}
\forall \mathbf{x} \in \Omega, \quad H ^ f \left ( \mathbf{x} \right ) \equiv \int _ {\mathbf{y} \in \Omega _ f} \delta \left ( \mathbf{x} - \mathbf{y} \right ) \mathrm{d} V
\label{eq:characteristic}
\end{equation}
where $\Omega$ denotes the computational domain and $\delta$ the multi-dimensional Dirac delta function. $H ^ f$ is not differentiable in the classical sense, and one should tread carefully not to interpolate or differentiate its surface- or volume-averaged values.







\subsection{Additional geometric information to restore accuracy} \label{sec:theanswer}

An intuitive idea to alleviate the interpolations in Eq.~\ref{eq:grad-v2} is to add new information where the volume (cell-centered and denoted $V$) and surface (face-centered and denoted $\left ( A _ \alpha \right )$) capacities were previously interpolated. These new quantities, referred to as second-kind capacities, complement the already used first-kind capacities $V$ and $\left ( A _ \alpha \right )$. Volume (face-centered and denoted $\left ( W _ \alpha \right )$) and surface (cell-centered and denoted $\left ( B _ \alpha \right )$) forms will be defined in Sec.~\ref{sec:notation} for arbitrary geometries.

These additional quantities yield the final gradient formula
\begin{equation}
\forall \alpha \in \left \{ x, y \right \}, \quad \operatorname{grad} _ \alpha \left ( T, D \right ) = \frac{1}{W _ \alpha} \left [ \frac{\delta B _ \alpha T}{\delta \xi _ \alpha} + \frac{\delta ( \overline{A _ \alpha} ^ \alpha - B _ \alpha ) D}{\delta \xi _ \alpha} - \overline{\frac{\delta A _ \alpha}{\delta \xi _ \alpha} D} ^ \alpha \right ]
\label{eq:gradD}
\end{equation}
 that supersedes $\operatorname{grad} _ \alpha ^ {\left ( \mathrm{v}1 \right )}$ and $\operatorname{grad} _ \alpha ^ {\left ( \mathrm{v}2 \right )}$.

\begin{figure}
    \centering
    \includegraphics[width=\textwidth]{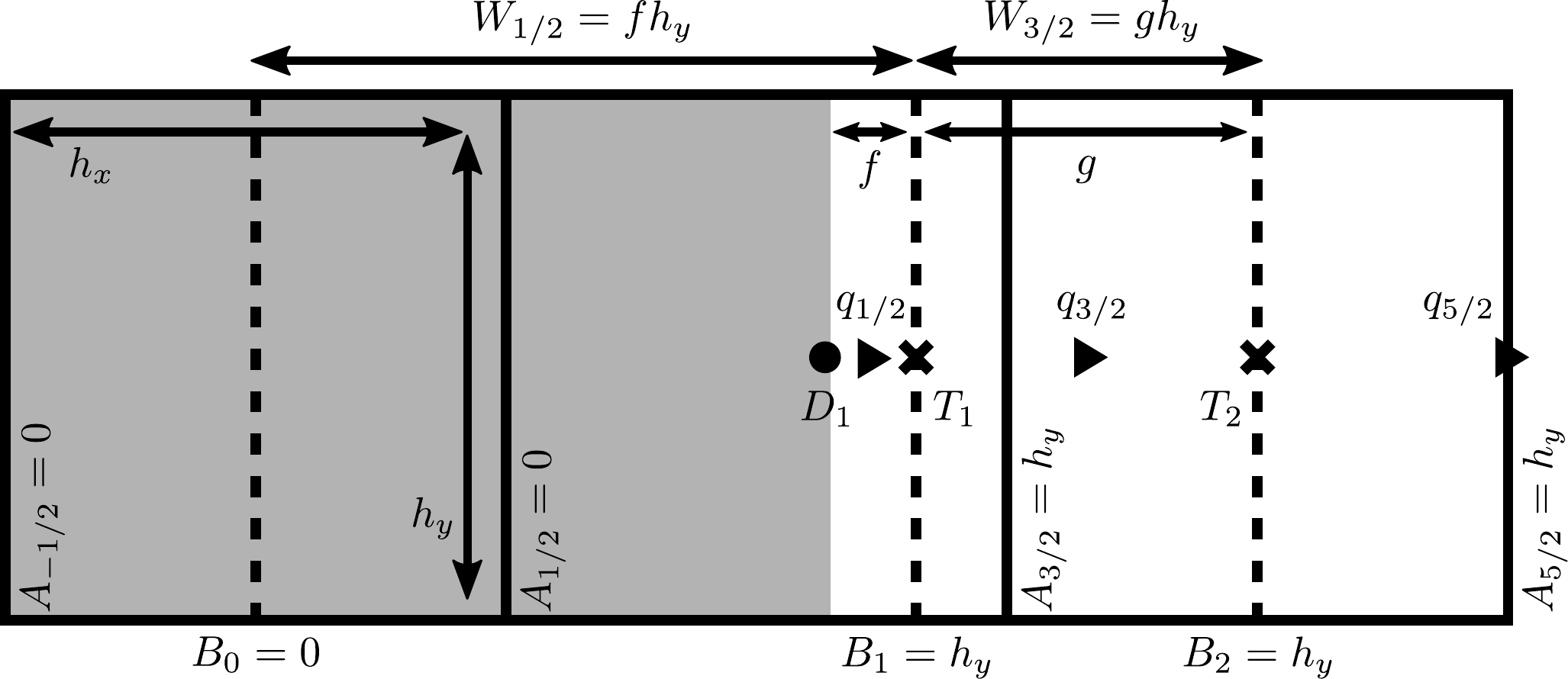}
    \caption{Enhanced geometric information restoring accuracy in mesh-aligned geometries.}
    \label{fig:orthoDoutbis}
\end{figure}

To show how the addition of the second-kind capacity restores first-order accuracy in the gradient computation, the configuration displayed in Fig.~\ref{fig:orthoDoutbis} is considered. Since only $x$ derivatives are considered, $A$ again will stands for $A _ x$, whereas $W$ and $B$ will respectively stand for $W _ x$ and $B _ x$. Bearing this in mind, the configuration under study is characterized by $V _ 0 = 0$, $V _ 1 = 2f h_y$ and $V _ 2 = h _ x h _ y$, $A _ {-1/2} = A _ {1/2} = 0$, $A _ {3/2} = A _ {5/2} = h _ y$, $B _ 0 = 0$, $B _ 1 = B _ 2 = h _ y$ and finally $W _ {-1 / 2} = 0$, $W _ {1/2} = f h _ y$, $W _ {3/2} = g h _ y$ and $W _ {5/2} = h _ x h _ y$. Using these expressions, Eq.~\ref{eq:gradD} simplifies to $Q _ {-1/2} = 0$ and
$$
Q _ {1 / 2} = \frac{T _ 1 - D _ 1}{f},
$$
$$
Q _ {3 / 2} = \frac{T _ 2 - T _ 1}{g}
$$
and
$$
Q _ {5 / 2} = \frac{T _ 3 - T _ 2}{h _ x}.
$$
$T _ 0$ does not appear since it is outside of the fluid domain, and the boundary condition ($D$) appears only in the faces adjacent to the boundary. The formulas obtained from Eq.~\ref{eq:gradD} are classical formulas since $f$, $g$ and $h _ x$ are the distances over which the differences $T _ 1 - D _ 1$, $T _ 2 - T _ 1$ and $T _ 3 - T _ 2$ are defined. Finally, in the fluid domain and away from the boundaries, Eq.~\ref{eq:gradD} simply reverts to the classical gradient formula
$$
Q _ {x, i + 1/2} = \frac{T _ {i + 1} - T _ i}{x _ {i + 1} - x _ i}.
$$




In fact, the addition of the second-kind capacities is also sufficient to define the (cell-centered) volume-weighted divergence operator, which consists of the sum of the contributions from $\forall \alpha \in \left \{ x, y \right \}$ where $N _ \alpha$ denotes the boundary value of $Q _ \alpha$. If one sets the divergence to the product of the volume $V$ with the local value of the source term $\sigma$ as in the original Poisson problem (Eq.~\ref{eq:laplacian}), the configuration displayed in Fig.~\ref{fig:orthoDoutbis} yields the trivial equation $0 = 0$ in the first cell, and
$$
h _ y \left ( Q _ {3 / 2} - N _ {1 / 2} \right ) = 2f h _ y \sigma _ 1,
$$
$$
h _ y \left ( Q _ {5 / 2} - Q _ {3 / 2} \right ) = h _ x h _ y \sigma _ 2
$$
in the rest. Again in the fluid domain and away from the boundary the classical formulas are obtained, given below
$$
h _ y \left ( Q _ {i + 1 / 2} - Q _ {i - 1 / 2} \right ) = h _ y \left ( x _ {i + 1 / 2} - x _ {i - 1 / 2} \right ) \sigma _ i.
$$
Finally, the unknown $\mathbf{N} = \left ( N _ \alpha \right )$ can be eliminated by substituting the gradient formula (Eq.~\ref{eq:gradD}) in the divergence formula defined below,
\begin{equation}
\operatorname{div} _ \alpha \left ( \underline{Q}, \underline{N} \right ) = \frac{\delta A _ \alpha Q _ \alpha}{\delta \xi _ \alpha} + \frac{\delta ( \overline{B} _ \alpha - A _ \alpha ) N _ \alpha}{\delta \xi _ \alpha} - \overline{\frac{\delta B _ \alpha}{\delta \xi} N _ \alpha} ^ \alpha.
\label{eq:divN}
\end{equation}
The boundary contribution (the last two terms in the right-hand side of Eq.~\ref{eq:divN}) are set to
$$
\sum _ \alpha \left [ \frac{\delta ( \overline{B} _ \alpha - A _ \alpha ) N _ \alpha}{\delta \xi _ \alpha} - \overline{\frac{\delta B _ \alpha}{\delta \xi} N _ \alpha} ^ \alpha \right ] = \sum _ \alpha \left [ \frac{\delta ( \overline{B} _ \alpha - A _ \alpha ) Q _ \alpha}{\delta \xi _ \alpha} - \overline{\frac{\delta B _ \alpha}{\delta \xi} Q _ \alpha} ^ \alpha \right ],
$$
which amount to identifying the heat flow through the boundary to the normal component of the temperature gradient.
In the configuration displayed in Fig.~\ref{fig:orthoDoutbis}, this yields one single non-trivial equation, $N _ 1 = Q _ 1$.

Putting it all together, the proposed gradient and divergence operators, defined for arbitrary boundary geometries in Eqs.~\ref{eq:gradD} and~\ref{eq:divN}, discretize the Poisson problem (Eq.~\ref{eq:laplacian}) in the configuration displayed in Fig.~\ref{fig:orthoDoutbis} as $0 = 0$,
$$
h _ y \left ( \frac{T _ 2 - T _ 1}{g} - \frac{T _ 1 - D}{f} \right ) = 2f h _ y \sigma _ 1
$$
and 
$$
h _ y \left ( \frac{T _ 3 - T _ 2}{h _ x} - \frac{T _ 2 - T _ 1}{g} \right ) = h _ x h _ y \sigma _ 2
$$
in the three cells displayed, while reverting to the classical formula
$$
h _ y \left ( \frac{T _ {i + 1} - T _ i}{x _ {i + 1} - x _ i} - \frac{T _ i - T _ {i-1}}{x _ i - x _ {i-1}} \right ) = h _ x h _ y \sigma _ i
$$
in the fluid domain away from the boundary.

As a consequence, formulas Eqs.~\ref{eq:gradD} and~\ref{eq:divN} can be interpreted as generalizations of the classical second-order formulas to accommodate the presence of arbitrary boundaries while preserving first-order accuracy in the presence of mesh-aligned cases.

\section{Definitions and notation} \label{sec:notation}

Before generalizing the methodology presented in Sec.~\ref{sec:methodology} to the discretization of the incompressible Navier-Stokes equations, this section clarifies the notation employed thus far, in particular the definition of volume and surface capacities of the first and second kinds for both cell- and face-centered quantities. The differentiation and interpolation operators are also recalled, and completed with the definition of the permanent product.

\subsection{Mesh and geometry input}

As far as the Cartesian mesh is concerned, a rectilinear mesh with $n _ x \times n _ y$ cells is defined by specifying the following sets of user-defined abscissas
$$
x _ {\sfrac{1}{2}} < x _ {\sfrac{3}{2}} < \cdots < x _ {n _ x + \sfrac{1}{2}}
$$
and
$$
y _ {\sfrac{1}{2}} < y _ {\sfrac{3}{2}} < \cdots < y _ {n _ y + \sfrac{1}{2}}.
$$
Importantly, the mesh need not be uniform. Any given cell $\Omega _ {ij}$, identified by a multi-index $ij$, $\left ( i, j \right ) \in \left \llbracket 1, n _ x \right \rrbracket \times \left \llbracket 1, n _ y \right \rrbracket$, corresponds to the set of points $\left ( x, y \right )$ that simultaneously satisfy $x _ {i - \sfrac{1}{2}} < x < x _ {i + \sfrac{1}{2}}$ and $y _ {j - \sfrac{1}{2}} < y < y _ {j + \sfrac{1}{2}}$.

\begin{figure}
     \centering
     \includegraphics[width=0.45\textwidth]{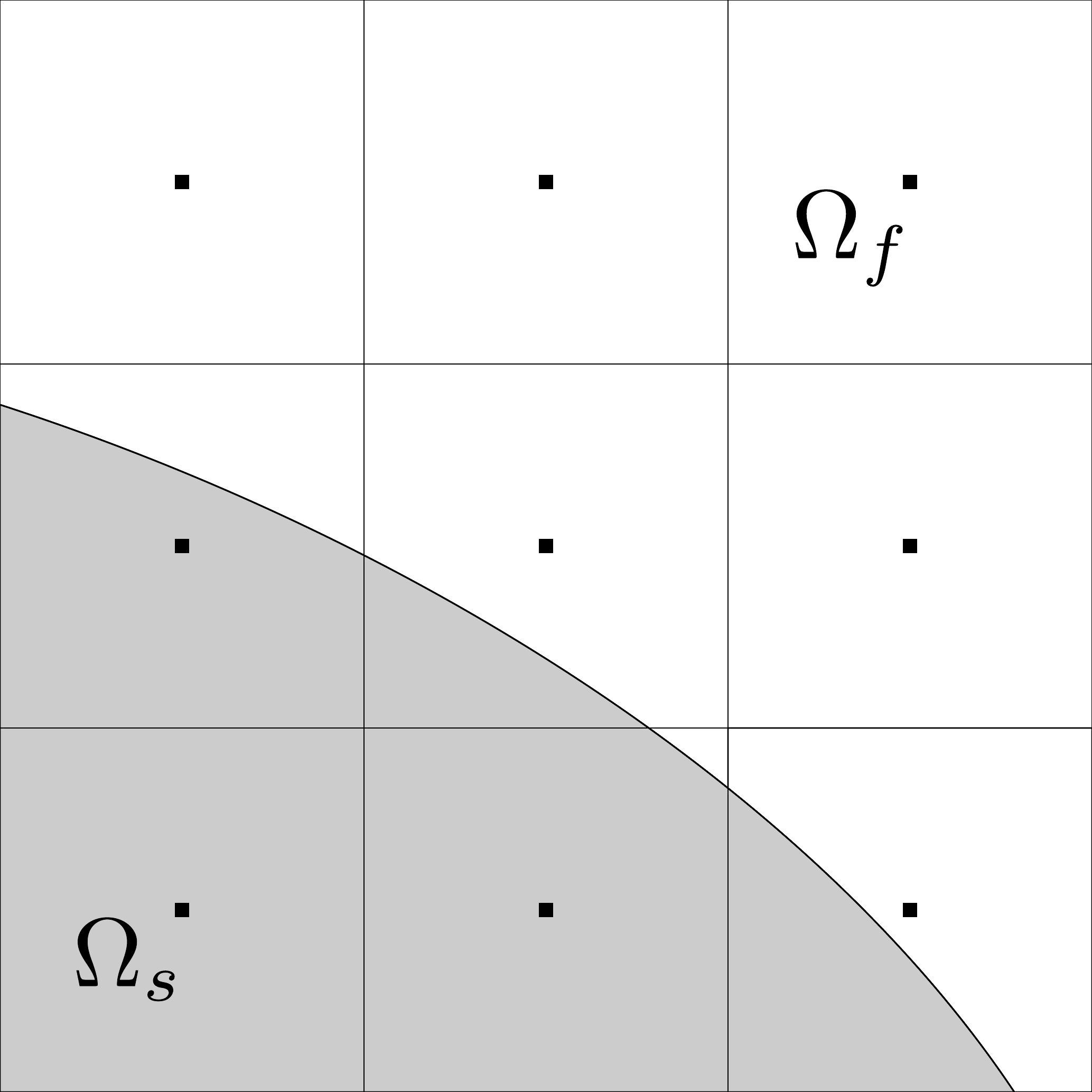}
     \caption{Intersection of the fluid domain ($\Omega _ f$) with Cartesian elements.}
     \label{fig:domains}
\end{figure}

Regarding the boundary description, there exists a wide range of techniques to define a fluid domain, such as simplicial meshes or Constructive Solid Geometry (CSG) primitives and operations. Implicit representations by means of a void fraction or distance function (Level Set) are also commonly used~\cite{Bloomenthal1997,Sethian1999}.
Regardless of the method employed, the assembly of the cut cell operators requires the computation of areas and volumes that correspond to the intersection of the fluid domain with Cartesian elements (faces or cells), as displayed in Fig.~\ref{fig:domains}.

In the proposed work, these computations are performed using either the Vofi library~\cite{Chierici2022} or a Marching Squares/Cubes algorithm~\cite{Lorensen1987}, both of which only require a signed distance function, readily available in the context of the Level Set method but which requires some implementation efforts in other input methods. This choice was made out of convenience, and other methods, such as ray tracing, can equally well work as placeholders.
Following the computation of the capacities, the geometry input is discarded.

\subsection{Capacities of the first kind}

Consider the Cartesian mesh displayed in Fig.~\ref{fig:domains}, partitioned into fluid ($\Omega ^ f$) and solid ($\Omega ^ s$) domains separated by a boundary ($\Gamma$). In a finite volume setting, the primary variables $\Phi _ {ij}$ consist of averages of any given continuous field $\left ( x, y \right ) \mapsto \phi \left ( x, y \right )$ over the intersection of the fluid domain with any given hexahedral cell, defined as follows
\begin{equation}
V _ {ij} \equiv \int _ {\Omega _ {ij}} \phi \left ( \mathbf{x} \right ) H ^ f \left ( \mathbf{x} \right ) \mathrm{d} ^ 2 \mathbf{x},
\label{eq:volumefirstkind}
\end{equation}
$$
\phi _ {ij} V _ {ij} \equiv \int _ {\Omega _ {ij}} \phi \left ( \mathbf{x} \right ) H ^ f \left ( \mathbf{x} \right ) \mathrm{d} ^ 2 \mathbf{x}
$$
where $H ^ f$ is the fluid characteristic function defined in Eq.~\ref{eq:characteristic}. The set $V \equiv \left ( V _ {ij} \right )$ is referred to as the volume capacities of the first kind.

\begin{figure}
     \centering
     \includegraphics[width=0.45\textwidth]{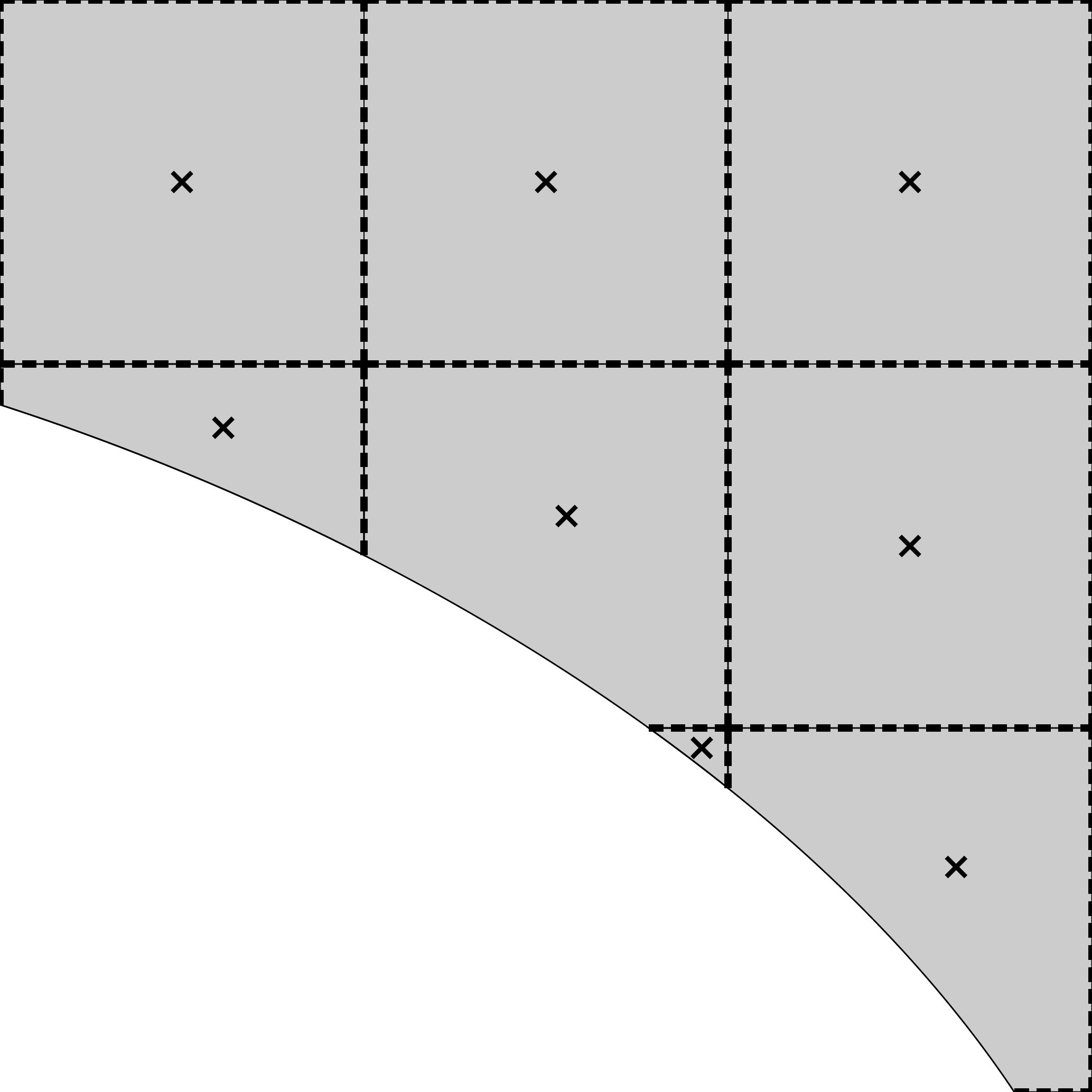}
     \caption{First kind capacities: $V$ (filled areas),  $A _ 1$ (dashed vertical lines), $A _ 2$ (dashed horizontal lines) and $X$ and $Y$ (crosses).}
     \label{fig:firstkind}
\end{figure}

When the field under consideration is linear, these averages coincide with the values at the fluid center of mass, displayed in Fig.~\ref{fig:firstkind}, defined as long as the cell is fully or partially occupied by the fluid. Although it does not appear explicitly in the cut cell operators, the coordinates of the fluid center of mass (displayed with crosses in Fig.~\ref{fig:firstkind}) are still required to define the second kind capacities, and are therefore temporarily stored. They are denoted as $X$ and $Y$ and defined for any cell $\Omega _ {ij}$ as
\begin{equation}
\left ( \begin{aligned} X _ {ij} \\  Y _ {ij} \end{aligned} \right ) V _ {ij} \equiv \int _ {\Omega _ {ij}} \left ( \begin{aligned} x \\ y \end{aligned} \right ) H ^ f \left ( \mathbf{x} \right ) \mathrm{d} ^ 2 \mathbf{x}.
\label{eq:centroid}
\end{equation}


The second step consists in computing the area of each of the faces wet by the fluid. Because the mesh is Cartesian, the faces adjacent to each cell are labelled based on the direction they are orthogonal to. These quantities, referred to as surface capacities, are staggered and are denoted as $\left ( A _ \alpha \right )$ ($\alpha \in \left \{ x, y \right \}$), and are defined as
\begin{equation}
A ^ x _ {i + \sfrac{1}{2}, j} \equiv \int _ {y _ {j - \sfrac{1}{2}}} ^ {y _ {j + \sfrac{1}{2}}} H ^ f \left ( x _ {i + \sfrac{1}{2}}, y \right ) \mathrm{d} y
\label{eq:surfacefirstkindx}
\end{equation}
and
\begin{equation}
A ^ y _ {i, j + \sfrac{1}{2}} \equiv \int _ {x _ {i - \sfrac{1}{2}}} ^ {x _ {i + \sfrac{1}{2}}} H ^ f \left ( x, y _ {j + \sfrac{1}{2}} \right ) \mathrm{d} x.
\label{eq:surfacefirstkindy}
\end{equation}

\subsection{Capacities of the second kind}

\begin{figure}
     \centering
     \begin{subfigure}[b]{0.45\textwidth}
         \centering
         \includegraphics[width=\textwidth]{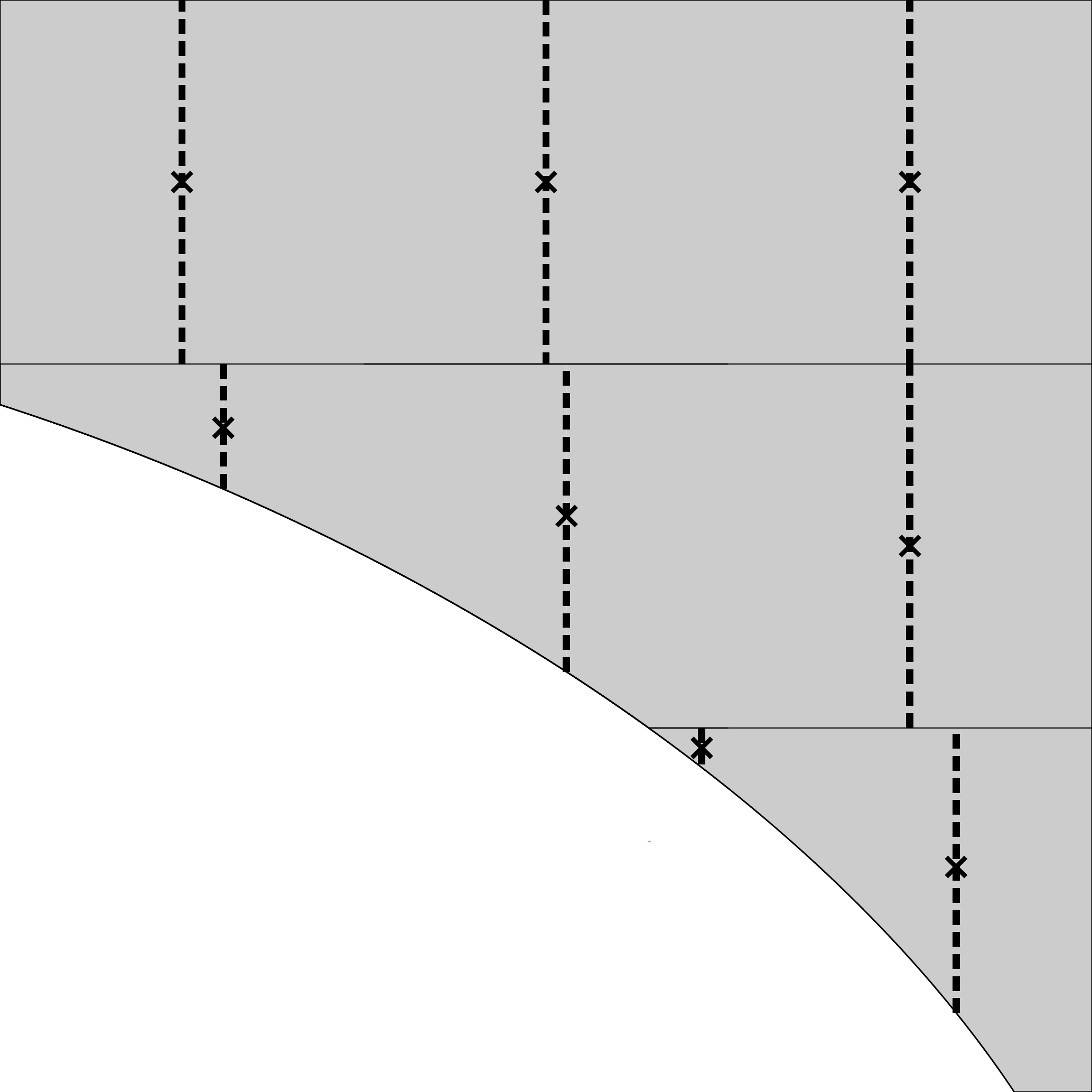}
         \caption{$W _ 1$ (filled areas),  $B _ 1$ (dashed vertical lines) and $X$ and $Y$ (crosses).}
         \label{fig:secondkindx}
     \end{subfigure}
     \hfill
     \begin{subfigure}[b]{0.45\textwidth}
         \centering
         \includegraphics[width=\textwidth]{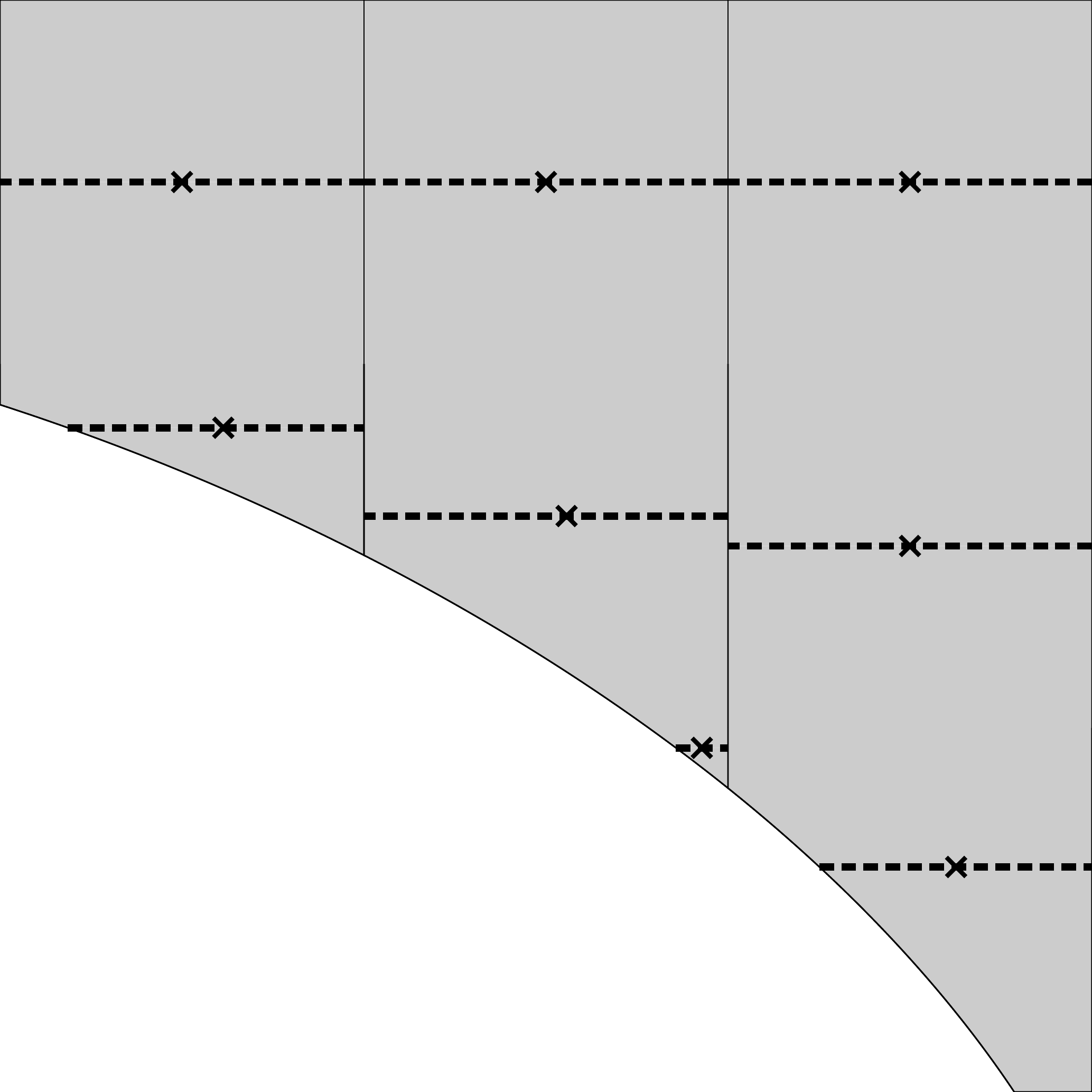}
         \caption{$W _ 2$ (filled areas),  $B _ 2$ (dashed horizontal lines) and $X$ and $Y$ (crosses).}
         \label{fig:secondkindy}
     \end{subfigure}
     \caption{Second kind capacities.}
     \label{fig:secondkind}
\end{figure}

The coordinates of the fluid center of mass are used as follows. For each direction, the volume information is enriched by measuring how much fluid lies between each center of mass. This yields as many sets of staggered volumes denoted as $\left ( W ^ \alpha \right )$, $\alpha \in \left \{ x, y \right \}$, defined as
\begin{equation}
W ^ x _ {i + \sfrac{1}{2},j} \equiv \int _ {y _ {j - \sfrac{1}{2}}} ^ {y _ {j + \sfrac{1}{2}}} \int _ {X _ {ij}} ^ {X _ {i+1,j}} H ^ f \left ( \mathbf{x} \right ) \mathrm{d} ^ 2 \mathbf{x}
\label{eq:volumesecondkindx}
\end{equation}
and
\begin{equation}
W ^ y _ {i, j + \sfrac{1}{2}} \equiv \int _ {Y _ {ij}} ^ {Y _ {i,j+1}} \int _ {x _ {i - \sfrac{1}{2}}} ^ {x _ {i + \sfrac{1}{2}}} H ^ f \left ( \mathbf{x} \right ) \mathrm{d} ^ 2 \mathbf{x}
\label{eq:volumesecondkindy}
\end{equation}
and referred to as volume capacities of the second kind. The capacities $W ^ x$ and $W ^ y$ are represented as colored areas in the configuration displayed in Fig.~\ref{fig:secondkindx} and~\ref{fig:secondkindy}, respectively.

Likewise, the area wet by the fluid for the mesh-aligned faces that intercept the fluid center of mass will be required in each cell. This yields an additional set of cell-centered quantities,
\begin{equation}
B ^ x _ {ij} = \int _ {y _ {j - \sfrac{1}{2}}} ^ {y _ {j + \sfrac{1}{2}}} H ^ f \left ( X _ {ij}, y \right ) \mathrm{d} y
\label{eq:surfacesecondkindx}
\end{equation}
and
\begin{equation}
B ^ y _ {ij} = \int _ {x _ {i - \sfrac{1}{2}}} ^ {x _ {i + \sfrac{1}{2}}} H ^ f \left ( x, Y _ {ij} \right ) \mathrm{d} x,
\label{eq:surfacesecondkindy}
\end{equation}
referred to a surface capacities of the second kind. The capacities $B ^ x$ and $B ^ y$ are represented as colored dashed lines in the configurations displayed in Fig.~\ref{fig:secondkindx} and~\ref{fig:secondkindy}, respectively.


\subsection{Staggering of the velocity components} \label{sec:staggering}

It will be shown that the only capacities required for the cell-centered quantities (the pressure field) are the surface capacities of the first kind
$$
( A ^ x _ {i + \sfrac{1}{2}, j} ) \quad \mathrm{and} \quad ( A ^ y _ {i, j + \sfrac{1}{2}} ).
$$

Considering the velocities however given the staggering of the $x$ and $y$ components, the computation of two additional sets of the first and second kind capacities are required, per velocity component. These computations are performed for the $x$ component by replacing $\left ( x _ {\sfrac{1}{2}}, \cdots, x _ {n_x + \sfrac{1}{2}} \right )$ abscissas by $\left ( x _ {0}, \cdots, x _ {n_x} \right )$, with half a grid spacing shift, and applying formulas of Eqs.~\ref{eq:volumefirstkind} and~\ref{eq:centroid}, Eqs.~\ref{eq:surfacefirstkindx} and~\ref{eq:surfacefirstkindy} and Eqs.~\ref{eq:volumesecondkindx}, \ref{eq:volumesecondkindy}, \ref{eq:surfacesecondkindx} and~\ref{eq:surfacesecondkindy} to compute the following first kind
$$
( V ^ x _ {i + \sfrac{1}{2}, j} ), \quad ( A ^ {xx} _ {i j} ) \quad \mathrm{and} \quad ( A ^ {xy} _ {i + \sfrac{1}{2}, j + \sfrac{1}{2}} )
$$
and second kind capacities
$$( W ^ {xx} _ {i j} ), \quad ( W ^ {xy} _ {i + \sfrac{1}{2}, j + \sfrac{1}{2}} ), \quad ( B ^ {xx} _ {i + \sfrac{1}{2}, j} ) \quad \mathrm{and} \quad ( B ^ {xy} _ {i + \sfrac{1}{2}, j} ).
$$

Likewise, abscissas $\left ( y _ {\sfrac{1}{2}}, \cdots, y _ {n_y + \sfrac{1}{2}} \right )$ are replaced by $\left ( y _ {0}, \cdots, y _ {n_y} \right )$, with half a grid spacing shift, to compute the capacities required for the $y$ component of the velocity field, yielding the following first kind
$$
( V ^ y _ {i, j+ \sfrac{1}{2}} ), \quad ( A ^ {yx} _ {i + \sfrac{1}{2}, j + \sfrac{1}{2}} ) \quad \mathrm{and} \quad ( A ^ {yy} _ {i j} )
$$
and second kind capacities
$$
( W ^ {yx} _ {i + \sfrac{1}{2}, j + \sfrac{1}{2}} ), \quad ( W ^ {yy} _ {i j} ), \quad  ( B ^ {yx} _ {i, j + \sfrac{1}{2}} ) \quad \mathrm{and} \quad ( B ^ {yy} _ {i, j + \sfrac{1}{2}} ).
$$

\section{Discretisation of the incompressible Navier-Stokes equations}

This section presents the proposed discretization of the incompressible Navier-Stokes equations for an isotropic Newtonian fluid
$$
\left \{ \begin{aligned}
\rho \left [ \frac{\partial \uline{u}}{\partial t} + \left ( \uline{u} \cdot \nabla \right ) \uline{u} \right ] & = -\nabla p + \nabla \cdot \left ( 2 \mu \uuline{s} \right ) + \rho \uline{g}, \\
\nabla \cdot \uline{u} & = 0
\end{aligned} \right .
$$
where $\uline{u}$ and $p$ respectively denote the fluid's velocity and pressure fields, $\rho$ its constant density and $\uline{g}$ the gravitational acceleration. Additionally, $\mu$ denotes the fluid's constant dynamic viscosity and
$$
\uuline{s} \equiv \frac{\nabla \uline{u} + \left ( \nabla \uline{u} \right ) ^ \top}{2}
$$
the strain-rate tensor. $P = \left ( P _ {ij} \right )$ represents the (cell-centered) pressure field, and
$$
\uline{U} = \left ( U _ x, U _ y \right ) = \left ( ( U ^ x _ {i + \sfrac{1}{2}, j} ), ( U ^ y _ {i, j + \sfrac{1}{2}} ) \right )
$$
the (staggered) Cartesian components of the velocity field. Finally, $\uline{D} = ( D _ x, D _ y )$ denotes the (staggered) boundary conditions to be applied on the velocity field.

\subsection{Velocity divergence and pressure gradient}

Let $\Omega _ {ij} ^ f = \Omega _ {ij} \cap \Omega ^ f$ denote the subset of $\Omega _ {ij}$ wet by the fluid, $\mathbf{u}$ the continuous fluid velocity field and $\mathbf{d}$ the boundary condition. Then, Stokes' divergence theorem
$$
\int _ {\Omega _ {ij} ^ f} \nabla \cdot \mathbf{u} = \int _ {\partial \Omega _ {ij} ^ f \setminus \Gamma} \mathbf{u} \cdot \mathbf{n} + \int _ {\partial \Omega _ {ij} ^ f \cap \Gamma} \mathbf{d} \cdot \mathbf{n}
$$
states that the volume integral of the velocity divergence matches the net volume fluxes, summed over the surfaces immersed in the fluid itself and adjacent to the boundary. The former term, referred to as homogeneous, quantifies the exchange of volume with the neighboring fluid elements, and the latter, referred to as heterogeneous, quantifies this exchange with the exterior domain through the boundary.

This decomposition is reflected at the discrete level by discretizing the volume-integrated velocity divergence as
\begin{equation}
\operatorname{cont} \left ( \uline{U}, \uline{D} \right ) \equiv \sum _ \alpha \left ( \frac{\delta A _ \alpha U _ \alpha}{\delta \xi _ \alpha} + \frac{\delta \left ( \overline{B _ \alpha} ^ \alpha - A _ \alpha \right ) D _ \alpha}{\delta \xi _ \alpha} - \overline{\frac{\delta B _ \alpha}{\delta \xi _ \alpha} D _ \alpha} ^ \alpha \right ).
\label{eq:divergence}
\end{equation}
The divergence free condition, then, is expressed as
$$
\operatorname{cont} \left ( \uline{U}, \uline{D} \right ) = 0
$$
and the (volume integrated) $\alpha$ component of the pressure gradient, a linear operator denoted as $\operatorname{pres} _ \alpha$, is simply defined as the negative transpose of the Jacobian of Eq.~\ref{eq:divergence} with respect to $U _ \alpha$, namely
\begin{equation}
\forall \alpha \in \left \{ x, y \right \}, \quad \frac{\partial \operatorname{pres} _ \alpha}{\partial P} = -\left ( \frac{\partial \operatorname{cont}}{\partial U _ \alpha} \right ) _ {U _ {\beta \ne \alpha}, \uline{D}} ^ \top
\label{eq:compatibility}
\end{equation}
which yields
$$
\forall \alpha \in \left \{ x, y \right \}, \quad \operatorname{pres} _ \alpha \left ( P \right ) \equiv A _ \alpha \frac{\delta P}{\delta \xi _ \alpha}.
$$
This construction is rooted in the geometric interpretation of the incompressible Navier-Stokes equations~\cite{Arnold}, which exposes the dual role of the pressure in imposing the divergence-free condition, and commonly used in both structured and unstructured settings~\cite{Cheny2010,Perot2011}.



\subsection{Strain-rate tensor}

The components of the diagonal element of the strain-rate tensor are cell-centered discrete counterparts of
$$
s _ {\alpha\alpha} = \frac{\partial u _ \alpha}{\partial x _ \alpha}, \quad \alpha \in \left \{ x, y \right \},
$$
defined based upon the gradient formula Eq.~\ref{eq:gradD}. First, the surface and volume capacities $\uline{W} = \left ( W _ \beta \right )$, $\uline{A} = \left ( A _ \beta \right )$ and $\uline{B} = \left ( B _ \beta \right )$ are replaced by those after shifting the mesh in half a grid spacing along direction $\alpha$ defined in Sec.~\ref{sec:staggering}, namely $\uline{W} ^ \alpha = \left ( W _ {\alpha\beta} \right )$, $\uline{A} ^ \alpha = \left ( A _ {\alpha\beta} \right )$ and $\uline{B} = \left ( B _ {\alpha\beta} \right )$. Second, the dependent field $T$ and the Dirichlet boundary condition $D$ are substituted with $U _ \alpha$ and $D _ \alpha$, respectively. This finally yields
\begin{multline}
\forall \alpha \in \left \{ x, y \right \}, \quad \operatorname{strain} _ {\alpha\alpha} \left ( \uline{U}, \uline{D} \right ) = \\
\frac{1}{W _ {\alpha\alpha}} \left [ \frac{\delta B _ {\alpha\alpha} U _ \alpha}{\delta \xi _ \alpha} + \frac{\delta \left ( \overline{A _ {\alpha\alpha}} ^ \alpha - B _ {\alpha\alpha} \right ) D _ \alpha}{\delta \xi _ \alpha} - \overline{\frac{\delta A _ {\alpha\alpha}}{\delta \xi _ \alpha} D _ \alpha} ^ \alpha \right ].
\label{eq:diag}
\end{multline}
This process is repeated for the components of the off-diagonal elements of the strain-rate tensor, defined in the continuous case as
$$
s _ {\alpha\beta} = \frac{1}{2} \left ( \frac{\partial u _ \alpha}{\partial x _ \beta} + \frac{\partial u _ \beta}{\partial x _ \alpha} \right ), \quad \alpha \ne \beta,
$$
and in the discrete case as the node-centered field
\begin{multline}
\forall \left ( \alpha, \beta \right ) \in \left \{ x, y \right \} ^ 2, \quad \alpha \ne \beta, \quad \operatorname{strain} _ {\alpha\beta} \left ( \uline{U}, \uline{D} \right ) = \\
\frac{1}{2W _ {\alpha\beta}} \left [ \frac{\delta B _ {\alpha\beta} U _ \alpha}{\delta \xi _ \beta} + \frac{\delta \left ( \overline{A _ {\alpha\beta}} ^ \beta - B _ {\alpha\beta} \right ) D _ \alpha}{\delta \xi _ \beta} - \overline{\frac{\delta A _ {\alpha\beta}}{\delta \xi _ \beta} D _ \alpha} ^ \beta \right ] \\
+ \frac{1}{2W _ {\beta\alpha}} \left [ \frac{\delta B _ {\beta\alpha} U _ \beta}{\delta \xi _ \alpha} + \frac{\delta \left ( \overline{A _ {\beta\alpha}} ^ \alpha - B _ {\beta\alpha} \right ) D _ \beta}{\delta \xi _ \alpha} - \overline{\frac{\delta A _ {\beta\alpha}}{\delta \xi _ \alpha} D _ \beta} ^ \alpha \right ].
\label{eq:offdiag}
\end{multline}
It should finally be noted that the latter formula (Eq.~\ref{eq:offdiag}) is also valid in the diagonal case ($\alpha = \beta$), in which case it simply reduces to Eq.~\ref{eq:diag}.

\subsection{Viscous transport term}

Prior to proceeding with the discretization of the viscous transport term, it should first be noted that, in the case where the second argument ($\uline{N}$) of the divergence operator (Eq.~\ref{eq:divN} summed over $\alpha$) matches the first argument ($\uline{Q}$), Eq.~\ref{eq:divN} may be simplified  using the identities presented by~\citet{Morinishi2010} as
\begin{equation}
\operatorname{div} \left ( \uline{Q}, \uline{Q} \right ) = \sum _ \beta B _ \beta \frac{\delta Q _ \beta}{\delta \xi _ \beta}.
\label{eq:divD}
\end{equation}
Therefore, the discretization of the viscous transport term, $\nabla \cdot \left ( 2 \mu s \right )$, is performed similarly to that of the strain-rate operator, by translating the definition of the capacities to yield
$$
\forall \alpha \in \left \{ x, y \right \}, \quad \operatorname{visc} _ \alpha \left ( \uuline{S} \right ) = \sum _ \beta B _ {\alpha \beta} \frac{\delta S _ {\alpha\beta}}{\delta \xi _ \beta}
$$
where $\uuline{S} = \left ( S _ {\alpha\beta} \right )$ is defined as a function of $\uline{U}$ and $\uline{D}$ by Eqs.~\ref{eq:diag} and~\ref{eq:offdiag}.

\subsection{Convective transport term}

The convective term in the momentum transport equation along $\alpha \in \left \{ x, y \right \}$ is rewritten in conservative form using the divergence-free condition,
$$
\left ( \uline{u} \cdot \nabla \right ) \uline{u} = \nabla \cdot \left ( \uline{u} \otimes \uline{u} \right ) - \nabla \cdot \uline{u} = \nabla \cdot \left ( \uline{u} \otimes \uline{u} \right )
$$
which in discrete form can be written as
\begin{multline}
\operatorname{conv} _ \alpha \left ( \uline{U}, \uline{U} ^ \dagger, \uline{D}, \uline{D} ^ \dagger \right ) = \\
\sum _ \beta \left \{ \frac{\delta \overline{A _ \beta U _ \beta} ^ \alpha \overline{U ^ \dagger _ \alpha} ^ \beta}{\delta \xi _ \beta} + \left [ \frac{\delta \overline{\left ( \overline{B _ \beta} ^ \beta - A _ \beta \right ) D _ \beta} ^ \alpha}{\delta \xi _ \beta} - \overline{\overline{\frac{\delta B _ \beta}{\delta \xi _ \beta} D _ \beta} ^ \beta} ^ \alpha \right ] \frac{U _ \alpha ^ \dagger + D ^ \dagger _ \alpha}{2} \right \}.
\label{eq:convective}
\end{multline}
This multilinear operator is typically evaluated at $\uline{U} ^ \dagger = \uline{U}$ and $\uline{D} ^ \dagger = \uline{D}$ but the distinction might bear significance, in the context of Picart linearisation for example where a distinction applies between $\uline{U}$ which is typically \emph{frozen} whereas $\uline{U} ^ \dagger$ is updated. 
This discretization can be considered as the generalisation of the centered scheme to the cut cell method, which can be demonstrated as follows. In the continuous case,
\begin{equation}
\forall \left ( \alpha, \beta \right \} \in \left \{ x, y \right \} ^ 2, \quad u _ \alpha \frac{\partial u _ \beta u _ \alpha}{\partial x _ \beta} = \frac{\partial u _ \beta u _ \alpha ^ 2 / 2}{\partial x _ \beta} + \frac{u _ \alpha ^ 2}{2} \frac{\partial u _ \beta}{\partial x_ \beta},
\label{eq:kecont}
\end{equation}
which, upon summation over $\alpha$, yields a similar equation for the specific kinetic energy $k \equiv \Vert \uline{u} \Vert ^ 2 / 2$, ultimately conserved in the inviscid limit. The proposed discretization of the convective transport term (Eq.~\ref{eq:convective}) preserves this property at the discrete level. Using the identities presented by~\citet{Morinishi2010}, it can be be shown that
\begin{equation}
\forall \left ( \alpha, \beta \right ) \in \left \{ x, y \right \} ^ 2, \quad U _ \alpha ^ \dagger \frac{\delta \overline{A _ \beta U _ \beta} ^ \alpha \overline{U ^ \dagger _ \alpha} ^ \beta}{\delta \xi _ \beta} = \frac{\delta \overline{A _ \beta U _ \beta} ^ \alpha \widetilde{U ^ \dagger _ \alpha U ^ \dagger _ \alpha} ^ \beta / 2}{\delta \xi _ \beta} + \frac{U ^ {\dagger 2} _ \alpha}{2} \frac{\delta \overline{A _ \beta U _ \beta} ^ \alpha}{\delta \xi _ \beta}
\label{eq:u-dot-qdm}
\end{equation}
where $\widetilde{\cdot}$ denotes the permanent product
\begin{equation}
\left . \widetilde{\phi \psi} ^ x \right \vert _ {i + \sfrac{1}{2}, j} = \frac{\phi_{i + 1, j} \psi_{i j} + \psi _ {i + 1, j} \phi _ {i j}}{2},
\label{eq:permanent}
\end{equation}
also introduced by~\citet{Morinishi2010} and easily extended to other dimensions and arrangements as previously done for differentiation and interpolation.
Eq.~\ref{eq:u-dot-qdm}, together with the continuity operator (Eq.~\ref{eq:divergence}), can be used to show that $\forall \alpha \in \left \{ x, y \right \}$
\begin{multline}
U _ \alpha ^ \dagger \operatorname{conv} _ \alpha \left ( \uline{U}, \uline{U} ^ \dagger, \uline{D}, \uline{D} ^ \dagger \right ) = \\
\sum _ \beta \left \{ \frac{\delta \overline{A _ \beta U _ \beta} ^ \alpha \widetilde{U ^ \dagger _ \alpha U ^ \dagger _ \alpha} ^ \beta / 2}{\delta \xi _ \beta} + \left [ \frac{\delta \overline{\left ( \overline{B _ \beta} ^ \beta - A _ \beta \right ) D _ \beta} ^ \alpha}{\delta \xi _ \beta} - \overline{\overline{\frac{\delta B _ \beta}{\delta \xi _ \beta} D _ \beta} ^ \beta} ^ \alpha \right ] \frac{U _ \alpha ^ \dagger D ^ \dagger _ \alpha}{2} \right \} \\
+ \frac{U ^ {\dagger 2} _ \alpha}{2} \overline{\operatorname{cont} \left ( \uline{U}, \uline{D} \right )} ^ \alpha.
\label{eq:kinetic}
\end{multline}
This identity can be interpolated in each direction $\alpha$, and summed over $\alpha$, to ultimately state the proposed discretization (Eq.~\ref{eq:convective}) conserves kinetic energy, in the sense that the rate of change of the discrete kinetic energy
$$
\operatorname{kinetic} \left ( \uline{U} ^ \dagger \right ) \equiv \sum _ \alpha \frac{1}{2} \overline{V _ \alpha U ^ \dagger _ \alpha U ^ \dagger _ \alpha} ^ \alpha
$$
is a result of an exchange with the neighboring fluid elements (first term in the right-hand side of Eq.~\ref{eq:kinetic}) and across the boundary (second term).

\subsection{Semi-discrete system}

The face-centered mass matrices appearing in front of the rate of change and body forces are diagonal with coefficients $\uline{V} = \left ( V _ \alpha \right )$ (the volume of the staggered control volumes, defined in Sec.~\ref{sec:staggering}) and are denoted as
$$
\forall \left \{ x, y \right \}, \quad \mathcal{M} _ \alpha \equiv \operatorname{diag} \left ( V _ \alpha \right ).
$$
Gathering all the terms, the proposed semi-discrete momentum equations then read ($\alpha \in \left \{ x, y \right \}$)
\begin{equation}
\rho \left [ \mathcal{M} _ \alpha \frac{\mathrm{d} U _ \alpha}{\mathrm{d} t} + \operatorname{conv} _ \alpha \left ( \uline{U}, \uline{U}, \uline{D}, \uline{D} \right ) \right ] =
-\operatorname{pres} _ \alpha \left ( P \right ) + \operatorname{visc} _ \alpha \left ( 2 \mu \uuline{S} \right ) + \rho \mathcal{M} _ \alpha \uline{g},
\label{eq:semidiscrete}
\end{equation}
with divergence-free condition
\begin{equation}
\operatorname{cont} \left ( \uline{U}, \uline{D} \right ) = 0.
\label{eq:continuity}
\end{equation}
The system is closed with the discrete strain-rate tensor $\uuline{S}$, defined as a function of $\uline{U}$ and $\uline{D}$ as follows,
\begin{equation}
\forall \left ( \alpha, \beta \right ) \in \left \{ x, y \right \} ^ 2, \quad S _ {\alpha\beta} = \operatorname{strain} _ {\alpha\beta} \left ( \uline{U}, \uline{D} \right )
\label{eq:constitutive}
\end{equation}
where the operators $\operatorname{strain} _ {\alpha\beta}$ are defined by Eqs.~\ref{eq:diag} and~\ref{eq:offdiag}.

All of the operators appearing in Eqs.~\ref{eq:semidiscrete}, \ref{eq:continuity} and~\ref{eq:constitutive} are linear in all dependent variables ($P$, $\uline{U}$ and $\uuline{S}$) and boundary condition $\uline{D}$ with the exception of the convective transport operators ($\left ( \operatorname{conv} _ \alpha \right )$ defined in Eq.~\ref{eq:convective}) which is quadratic when evaluated at $\uline{U} ^ \dagger = \uline{U}$ and $\uline{D} ^ \dagger = \uline{D}$.


\subsection{Projection method}

The discretization of the aforementioned incompressible Navier-Stokes equations results in a saddle point system of equations \cite{Benzi2005}, sometimes also called Karush-Kuhn-Tucker (KKT) system~\cite{Nocedal2006} in optimization. A wide range of algorithms have been devised to efficiently solve saddle point systems (or approximation thereof). In the field of fluid mechanics, a common approach is the fractional step method~\cite{Chorin1968}. In the present work, the method referred to as projection method II (PmII) by~\citet{Brown2001}, which ensures a second order discretization of the equations, is employed.

In this projection method, the convective term is discretized using the explicit second-order Adams-Bashforth scheme and the viscous term is discretized using the implicit Crank-Nicolson scheme. The first step of the method consists of obtaining an intermediate velocity field $\uline{U} ^ \star$ by solving
\begin{multline}
\rho \mathcal{M} _ \alpha \frac{U _ \alpha ^ \star - U _ \alpha ^ n}{\tau} + \frac{3 \rho}{2} \operatorname{conv} _ \alpha \left ( \uline{U} ^ n, \uline{U} ^ n, \uline{D} ^ n, \uline{D} ^ n \right ) \\
- \frac{\rho}{2} \operatorname{conv} _ \alpha \left ( \uline{U} ^ {n - 1}, \uline{U} ^ {n - 1}, \uline{D} ^ {n - 1}, \uline{D} ^ {n - 1} \right ) = -\operatorname{pres} _ \alpha \left ( P ^ {n - 1 / 2} \right ) \\
+ \operatorname{visc} _ \alpha \left ( \mu \uuline{S} ^ \star \right ) + \operatorname{visc} _ \alpha \left ( \mu \uuline{S} ^ n \right ) + \rho \mathcal{M} _ \alpha \uline{g},
\label{eq:pm_prediction}
\end{multline}
where $\tau$ denotes the time step and the superscript $n$ the iteration number. The boundary conditions applicable to $\uline{U} ^ \star$ (the predicted velocity field) and used in $\uuline{S} ^ \star$ are those of the velocity field at the next time step ($\uline{D} ^ \star = \uline{D} ^ {n+1}$)
$$
\forall \left ( \alpha, \beta \right ), \quad
S ^ n _ {\alpha \beta} = \operatorname{strain} _ {\alpha \beta} \left ( \uline{U} ^ n, \uline{D} ^ n \right )
\quad \mathrm{and} \quad
S ^ \star _ {\alpha \beta} = \operatorname{strain} _ {\alpha \beta} \left ( \uline{U} ^ \star, \uline{D} ^ {n + 1} \right ).
$$

In the projection step, the velocity field is updated by projecting $\uline{U} ^ \star$ using the intermediate pressure field $\Phi ^ {n + 1}$, which is obtained by solving the following Poisson equation
\begin{equation}
\tau \operatorname{cont} \left ( \uline{\operatorname{pres}}  \left ( \Phi ^ {n + 1} \right ), \uline{0} \right ) = \operatorname{cont} \left ( \uline{U} ^ \star, \uline{D} ^ {n + 1} \right ),
\label{eq:pm_poisson}
\end{equation}
with a homogeneous Neumann boundary conditions being used for the intermediate pressure ($\uline{0}$). The velocity field is ultimately corrected as
\begin{equation}
U _ \alpha ^ {n + 1} = U ^ \star _ \alpha - \tau \operatorname{pres} _ \alpha \left ( \Phi ^ {n + 1} \right ).
\label{eq:pm_projection}
\end{equation}

The pressure is finally updated as
\begin{equation}
P ^ {n + 1 / 2} = P ^ {n - 1 / 2} + \Phi ^ {n + 1} - \frac{\tau \mu}{2 \rho} \operatorname{cont} \left ( \uline{\operatorname{pres}} \left ( \Phi ^ {n + 1} \right ), \uline{0} \right ),
\label{eq:pm_update}
\end{equation}
where the last term ensures the second order accuracy of the pressure field.

Thus far, only Dirichlet boundary conditions for the velocity field have been considered, which are paired with homogeneous boundary conditions for the pressure in the projection step. Cases will be considered in the following section where Neumann boundary condition are required along the outflow boundaries. Along their vicinity, a Dirichlet boundary condition for the pressure is employed in order to uphold the compatibility equation~\ref{eq:compatibility}.

%
Finally, the use of periodic and/or Neumann boundary conditions gives rise to a rank deficiency in the Laplacian operator. This results in the pressure field being known up to a constant. This knowledge is exploited in the iterative solution of the Poisson equation by projecting the updates in the space of zero-mean solutions.

\section{Results}

Two canonical test cases are presented to validate the methodology and showcase that the proposed cut cell method is able to accommodate geometries of any shape.

\subsection{Flow around a cylinder}

\begin{figure}
\begin{floatrow}
\ffigbox{
  \includegraphics[width=0.35\textwidth]{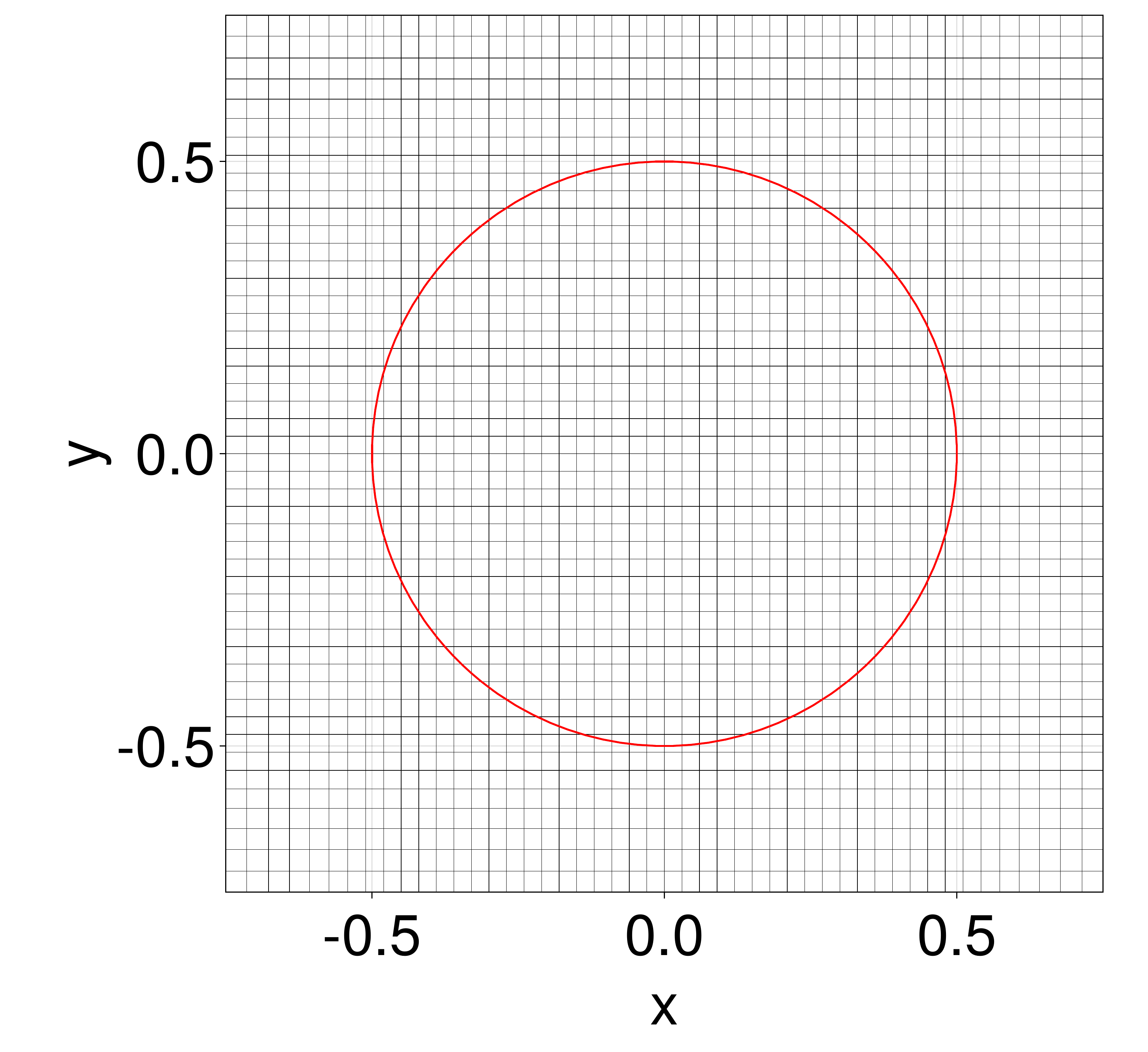}
}{
  \caption{Close-up view of grid G1.}
  \label{fig:g1}
}
\capbtabbox{
  \begin{tabular}{l l l l}
        \hline
        Grid & $n _ x \times n _ y$ & $\Delta x _ \mathrm{min}$ & $\Delta x _ \mathrm{max}$  \\
        \hline
        G1 & $320 \times 200$ & 0.06 & 0.2 \\
        G2 & $600 \times 350$ & 0.03 & 0.1 \\
        G3 & $1150 \times 500$ & 0.015 & 0.075 \\
        \hline
    \end{tabular}
}{
  \caption{Grids parameters for the cylinder.}
    \label{tab:grids_cylinder}
}
\end{floatrow}
\end{figure}

Viscous flow around a cylinder at $\operatorname{Re} = 100$  is used to test the accuracy of the proposed method. Three different grids have been tested with a domain size $\mathcal{D} = \left [ -15 , 30 \right ] \times \left [ -15 , 15 \right ]$ with varying resolutions, labelled G1 (coarsest) to G3 (finest), in order to assess the accuracy of the method in a canonical configuration and to highlight its convergence properties. Fig.~\ref{fig:g1} shows a close-up view of the grid G1, whereas the number of points in each direction and the minimum and maximum cell size of each grid are shown in Tab.~\ref{tab:grids_cylinder}.

Dirichlet boundary condition is applied on the left border of the domain on the velocity field whereas homogeneous Neumann boundary conditions are applied on the bottom, right and top borders as outflow boundary conditions. On the pressure field, homogeneous Neumann is applied on the left border and Dirichlet on the bottom, right and top borders. A no-slip Dirichlet boundary condition is used at the wall for the velocity and homogeneous Neumann for the pressure. The CFL number is set to 0.5 in all the simulations. The horizontal and vertical components of the velocity field are initialized to $1$ and $0$, respectively. The simulations are advanced $200$ time units in order to reach the periodic state.

\begin{figure}
    \centering
    \includegraphics[width=0.6\textwidth]{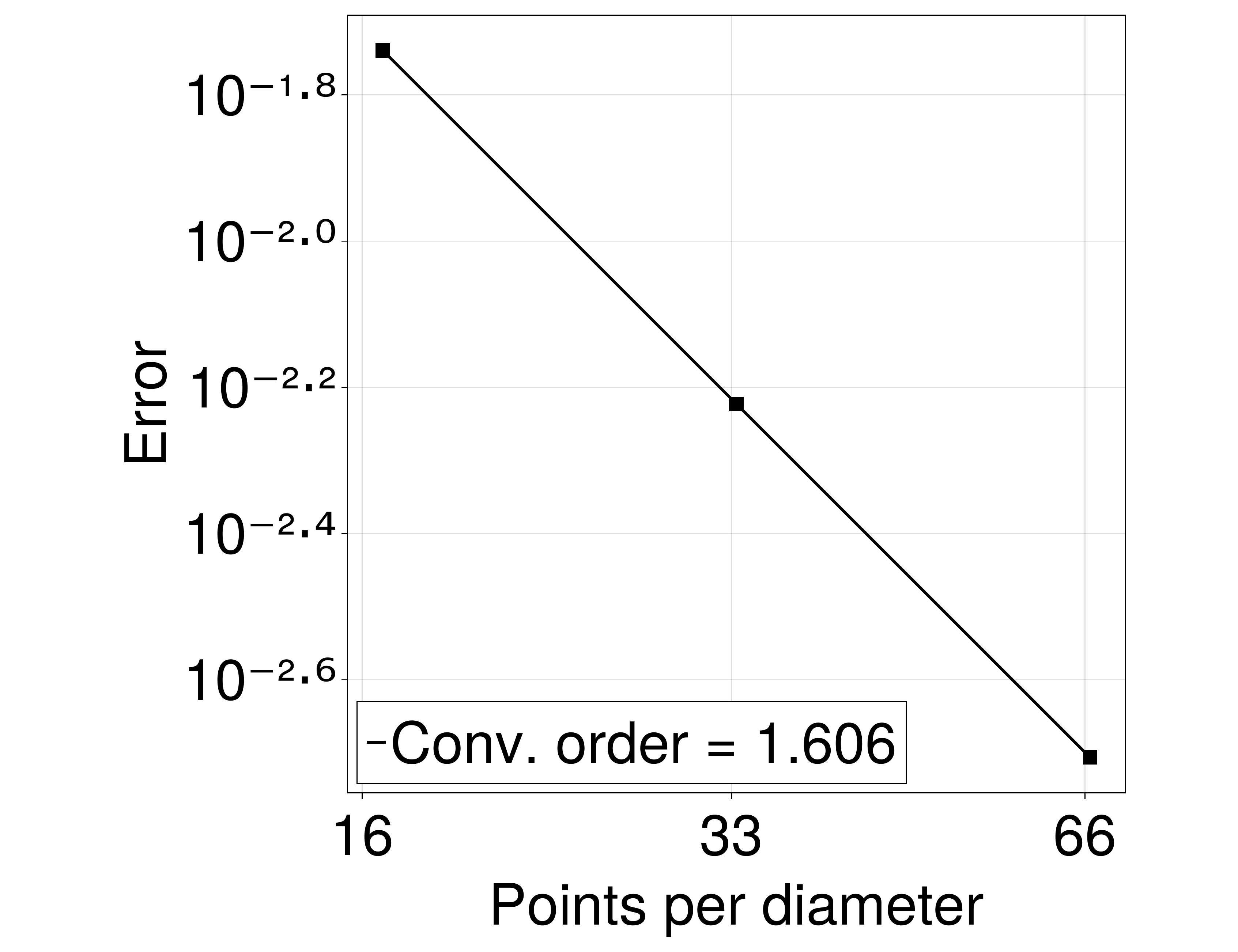}
    \caption{Convergence of $\overline{C _ d}$.}
    \label{fig:convergence}
\end{figure}

Fig.~ \ref{fig:convergence} depicts the error and the order of convergence of the proposed methodology by measuring the error as the difference in the mean drag coefficient between the values obtained using grids G1 and G2 and the value obtained using grid G3, which is used as reference. A convergence rate of 1.606 is observed. The results obtained for the Strouhal number ($\operatorname{St}$), the root mean square lift coefficient (r.m.s. $C _ l$) and the drag coefficient ($C _ d$) are presented in Tab.~\ref{tab:results_cylinder} for the finest grid G3 and compared with several reference solutions, showing a good agreement.

\begin{table}
    \centering
    \begin{tabular}{l l l l}
        \hline
         & $\operatorname{St}$ & r.m.s. $C _ l$ & $C _ d$ \\
        \hline
        G3 & 0.167 & 0.251 & $1.370 \pm 0.008$ \\
        Norberg \cite{Norberg2003} & 0.164 & 0.265 & - \\
        Henderson \cite{Henderson1997} & 0.164 & - & 1.350 \\
        He \textit{et al.} \cite{He2000} & 0.167 & - & 1.353 \\
        Linnick and Fasel \cite{Linnick2005} & 0.166 & - & $1.38 \pm 0.009$ \\
        \hline
    \end{tabular}
    \caption{Comparison of Strouhal number, r.m.s. lift coefficient and drag coefficient for the cylinder case at $\operatorname{Re} = 100$.}
    \label{tab:results_cylinder}
\end{table}

\begin{figure}
     \centering
     \begin{subfigure}[b]{0.45\textwidth}
        \centering
        \includegraphics[width=\textwidth]{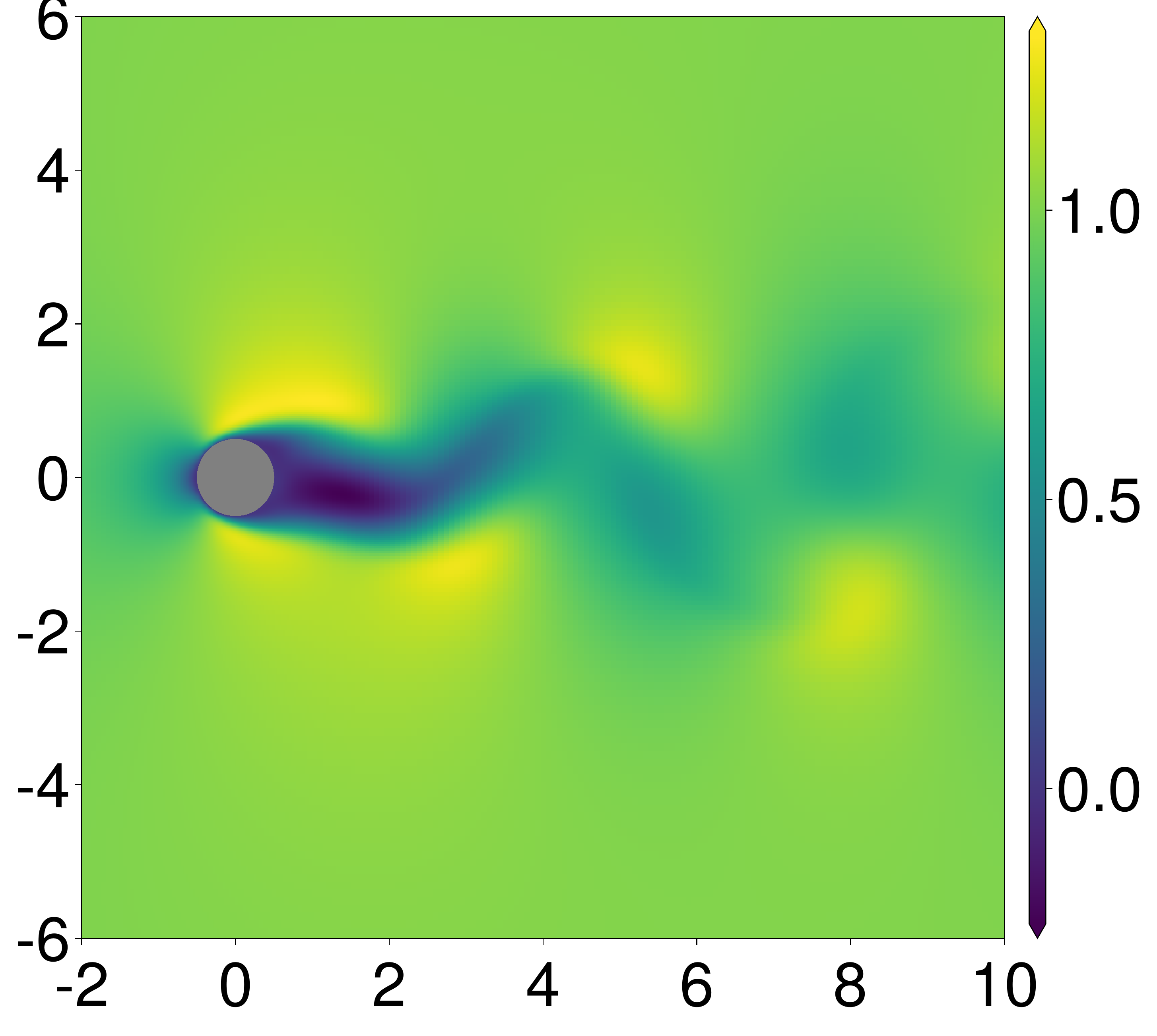}
        \caption{Horizontal velocity component.}
     \end{subfigure}
     \hfill
     \begin{subfigure}[b]{0.45\textwidth}
        \centering
        \includegraphics[width=\textwidth]{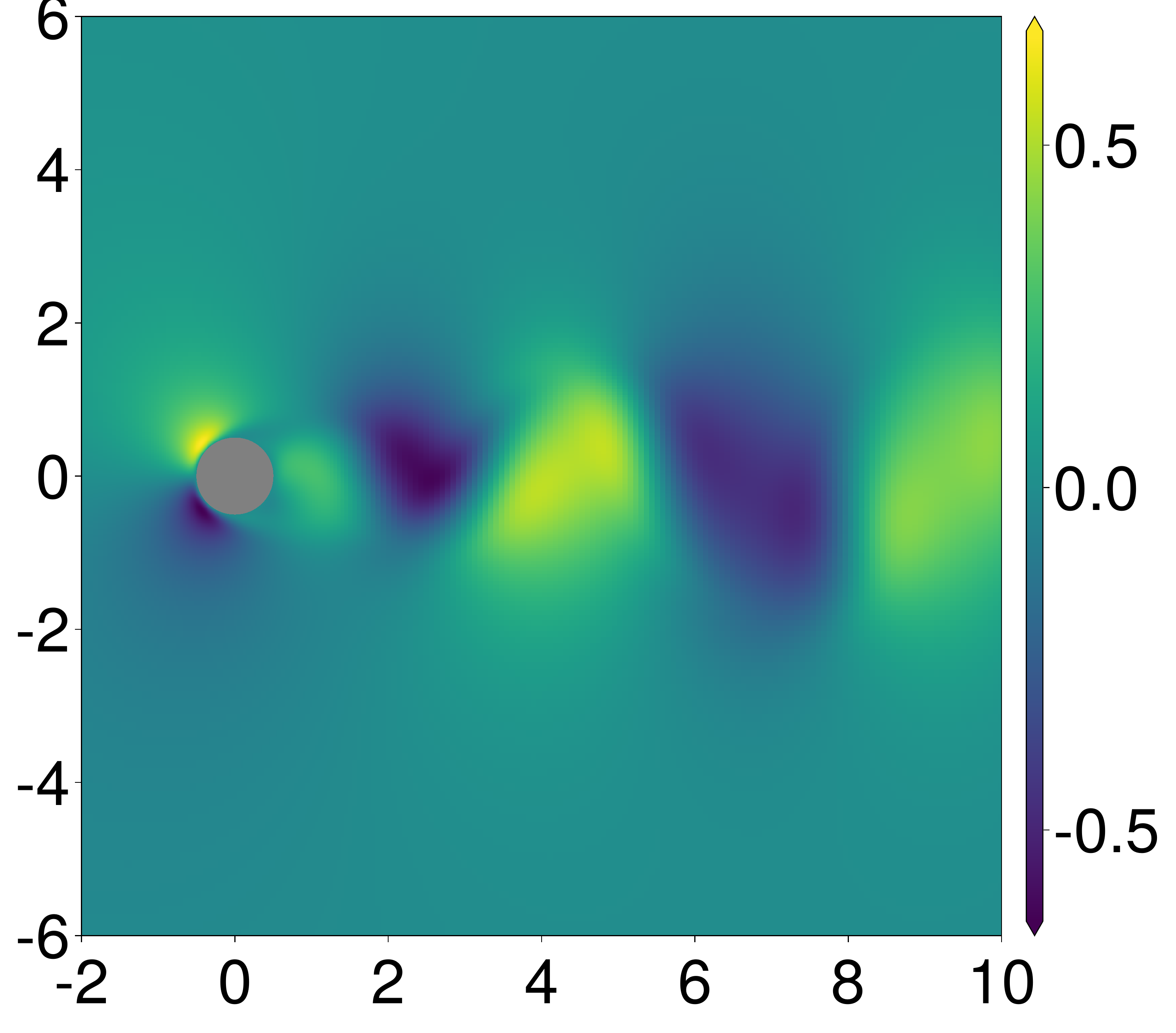}
        \caption{Vertical velocity component.}
     \end{subfigure}
     \caption{Colormap of the velocity components using grid G4 at $t = 200$.}
     \label{fig:vel_cylinder}
\end{figure}

A snapshot of the streamwise and vertical velocity fields at $t = 200$ is displayed in Fig.~\ref{fig:vel_cylinder} showing the expected von Kármán vortex street.

\subsection{Flow around an airfoil}

\begin{figure}
    \centering
    \includegraphics[width=\textwidth]{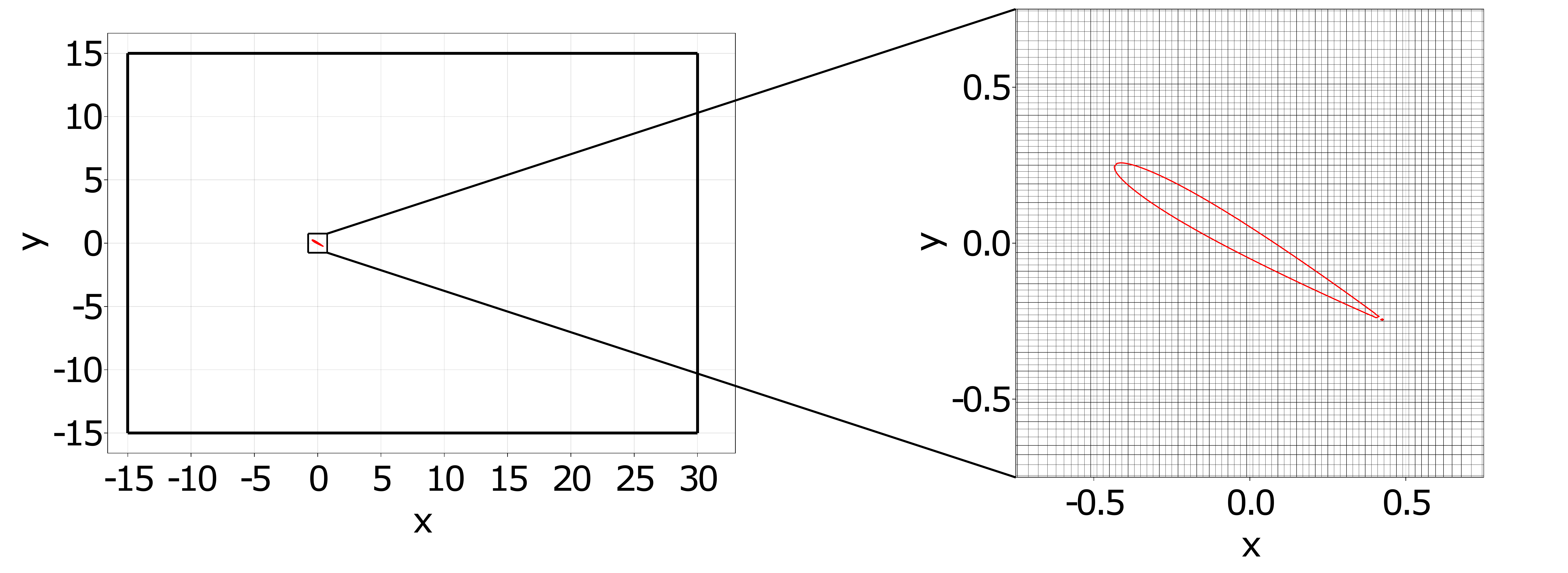}
    \caption{Overall and close-up views of the grid used for the NACA 0010 airfoil.}
    \label{fig:grid_airfoil}
\end{figure}

The flow around the symmetric NACA 0010 airfoil at $\operatorname{Re} = 500$ and an angle of attack $\alpha = 30 ^ \circ$ is also simulated and compared with a reference solution~\cite{Rossi2018}. In this case, a single grid has been used with a domain size $\mathcal{D} = [-15, 30] \times [-15, 15]$ using $1200 \times 600$ grid points, with a minimum cell size of $\Delta x _ \mathrm{min} = 0.01$ and a maximum cell size of $\Delta x _ \mathrm{max} = 0.075$. Fig.~\ref{fig:grid_airfoil} displays a general and a close-up view of the grid around the airfoil plotting the grid lines every two cells for the sake of clarity. As in the cylinder case, the horizontal component of the velocity is initialized to $1$, and the vertical component to $0$. The simulation is advanced $80$ time units until the periodic stated is reached and the same set of boundary conditions as those of the previous case are applied. The CFL number is set to $0.25$ in this case.

\begin{table}
    \centering
    \begin{tabular}{l l l l}
        \hline
         & $\operatorname{St}$ & $\overline{C _ l}$ & $\overline{C _ d}$ \\
        \hline
        Present & $0.36$ & $1.1$ & $0.77$ \\
        Rossi \textit{et al.}~\cite{Rossi2018} & $0.34$ & $1.1$ & $0.75$ \\
        \hline
    \end{tabular}
    \caption{Comparison of the Strouhal number, mean lift coefficient and mean drag coefficient for NACA 0010 at $\operatorname[Re] = 500$ and $\alpha = 30 ^ \circ$.}
    \label{tab:results_airfoil}
\end{table}

\begin{figure}
     \centering
     \begin{subfigure}[b]{0.45\textwidth}
        \centering
        \includegraphics[width=\textwidth]{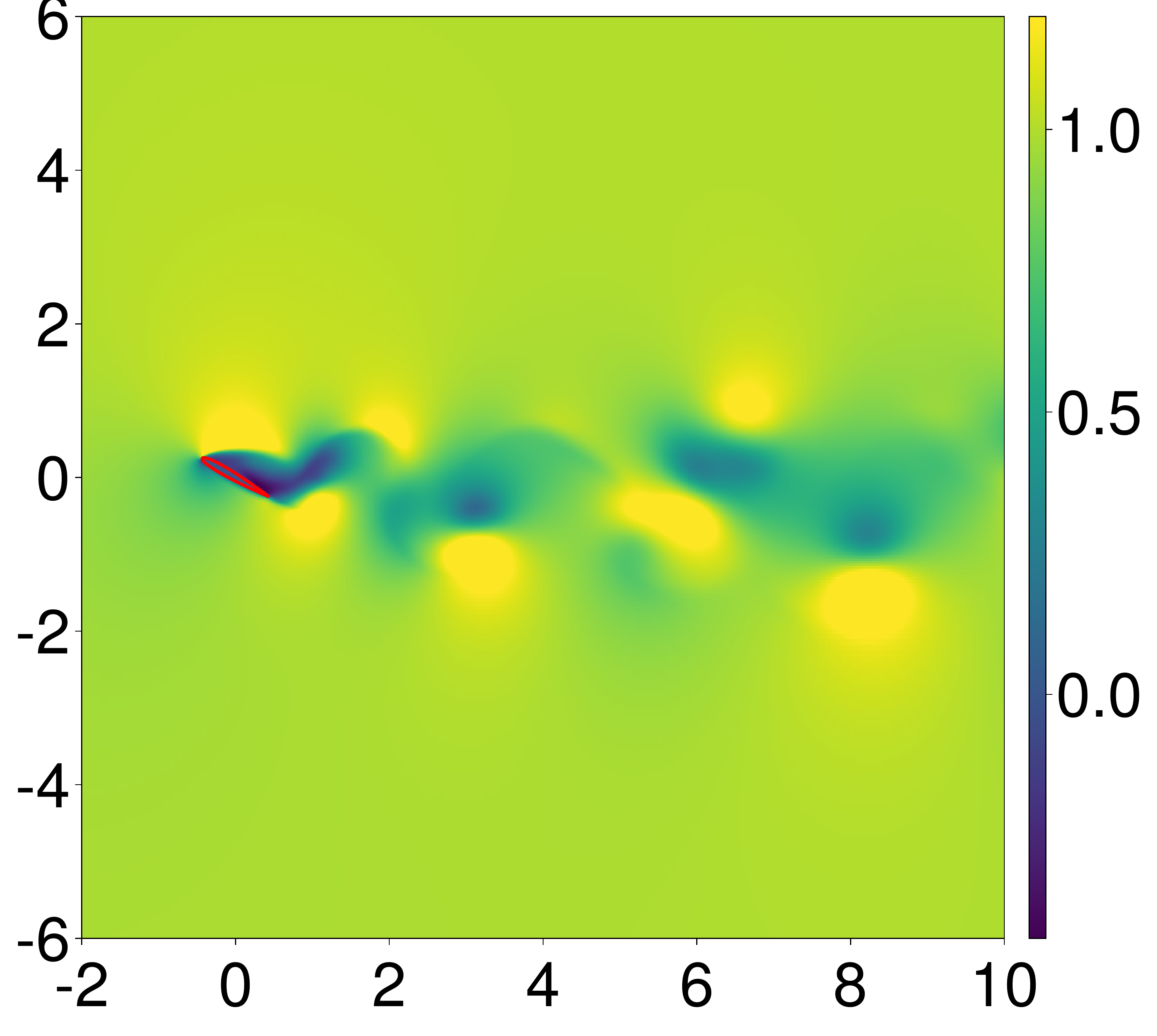}
        \caption{Horizontal velocity component.}
     \end{subfigure}
     \hfill
     \begin{subfigure}[b]{0.45\textwidth}
        \centering
        \includegraphics[width=\textwidth]{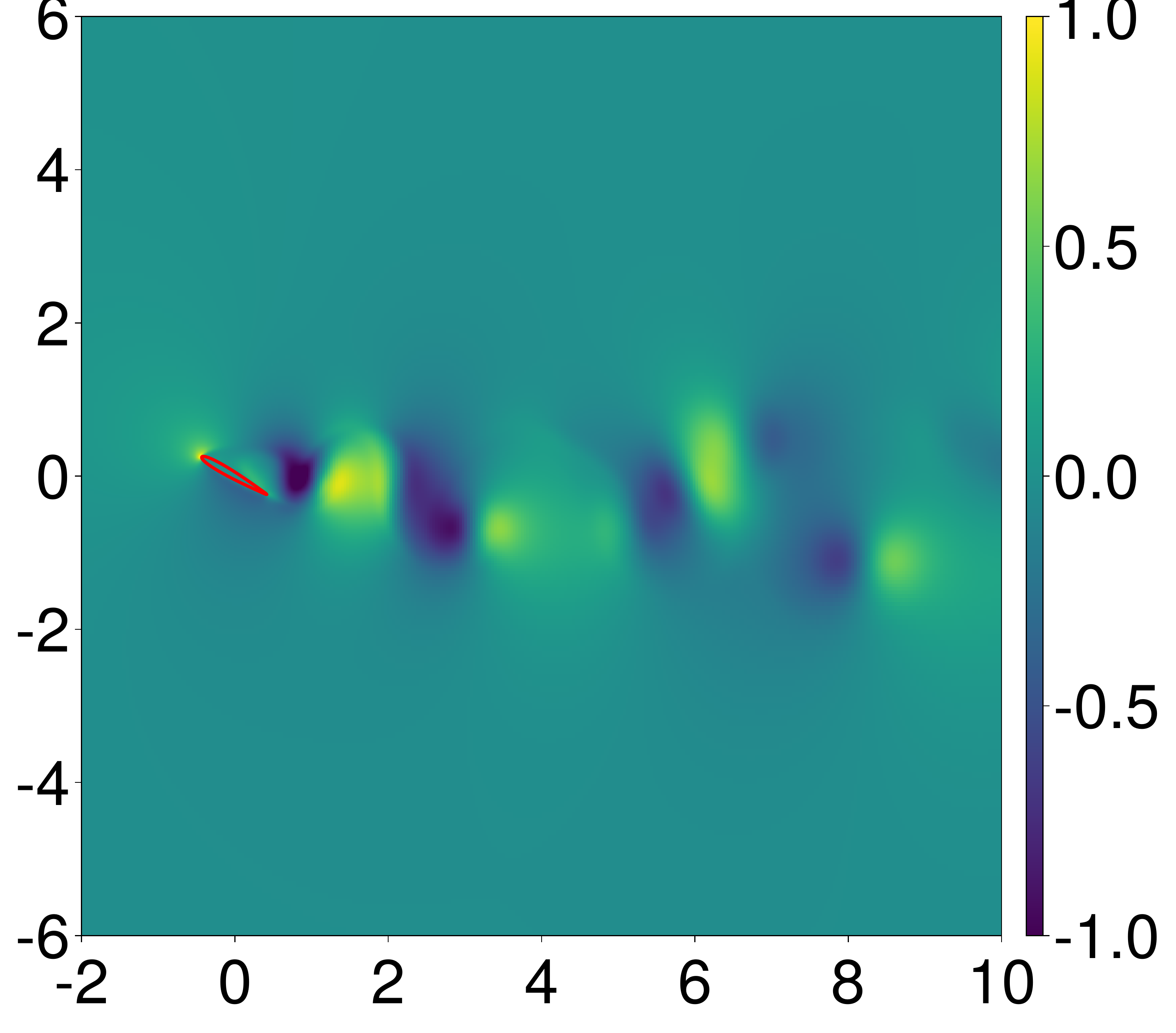}
        \caption{Vertical velocity component.}
     \end{subfigure}
     \caption{Colormaps of the velocity components around the airfoil at $\operatorname{Re} = 500$.}
     \label{fig:vel_airfoil}
\end{figure}

Fig.~\ref{fig:vel_airfoil} shows the velocity components at the last time step of the simulation, where the wake displays alternating vortex pairs being shed. One vortex pair is in vertical ascent while the other pair moves downstream following a descending path. This double vortex pair generates a double wake structure downstream of the airfoil.

\section{Conclusion}

The proposed cut cell methodology relies on Morninishi's discrete calculus to formulate discrete operators for the solution of the incompressible Navier-Stokes equations on staggered Cartesian grids in arbitrarily-shaped domains. Emphasis is set on both accuracy and structural properties of the first- and second-order operators. The geometric information is encapsulated in a set of surface and volume moments, designed to preserve constant states, recover classical formulas away from the boundary and in the vicinity of mesh-aligned boundaries, and retain a nearest-neighbor stencil. By construction, the spatial operators conserve volume and linear momenta locally and globally as well as kinetic energy in the inviscid limit. The method is shown to perform well in canonical two-dimensional flow configurations. Future work includes the generalisation to more complex boundary conditions as well as the replacement of the segregated approximation by a monolithic pressure-velocity solver.

\section*{Acknowledgements}

V.~{Le Chenadec} gratefully acknowledges Drs.~Y.Y.~Bay and A.~Fikl's valuable help in implementing an initial version of the cut-cell discretisation for the Navier-Stokes equations, described in Sec.~\ref{sec:bay} and documented elsewhere~\cite{Bay}. The work of A.~Quir{\'o}s~Rodr{\'i}guez and T.~Fullana was performed with the financial support from the ED SMAER and the ISCD at Sorbonne Universit\'{e}.

\end{document}